\documentclass[preprint,12pt]{elsarticle}

\usepackage[T1]{fontenc}

\usepackage[utf8]{inputenc}
\usepackage{lineno}
\modulolinenumbers[5]

\usepackage{amssymb}
\usepackage{amsmath}
\usepackage{pifont}
\newcommand{\cmark}{\ding{51}}%
\newcommand{\xmark}{\ding{55}}%
\usepackage{enumitem}
\usepackage{flafter}

\usepackage{graphicx}      % include this line if your document contains figures
\usepackage[numbers]{natbib}        % required for bibliography
\makeatletter
\patchcmd{\NAT@test}{\else \NAT@nm }{\else \NAT@nmfmt{\NAT@nm}}{}{}
\makeatother
\usepackage[colorlinks=true]{hyperref}
\usepackage{xurl}
\hypersetup{
    colorlinks=true,
    linkcolor=black,
    filecolor=magenta,
    urlcolor=blue,
    citecolor=black,
}
\usepackage{wrapfig}
\usepackage{graphicx}
\usepackage{tabularx}
\usepackage{booktabs}
\usepackage{setspace}

\usepackage{xcolor}
\newcolumntype{R}{>{\raggedleft\arraybackslash}X}
\newcolumntype{L}{>{\raggedright\arraybackslash}X}

\usepackage[capitalise, noabbrev]{cleveref} 
\crefname{appendix}{}{}
\usepackage{comment}

\usepackage{acro}
\acsetup{single={1},	
	barriers/use=true,
	barriers/reset=true,
	barriers/single=true}
\DeclareAcronym{AI}{short = AI,
	long = artificial intelligence}

\DeclareAcronym{RL}{short = RL,
	long = reinforcement learning,
	short-indefinite = an}
	
\DeclareAcronym{ML}{short = ML,
	long = machine learning,
	short-indefinite = an}

\DeclareAcronym{CCA}{short = CCA,
	long = canonical correlation analysis}	

\DeclareAcronym{FA}{short = FA,
	long = factorial analysis}
	
\DeclareAcronym{GMM}{short = GMM,
	long = Gaussian mixture model}	

\DeclareAcronym{ICA}{short = ICA,
	long = independent component analysis}	

\DeclareAcronym{LARS}{short = LARS,
	long = least-angle regression}	
			
\DeclareAcronym{LASSO}{short = LASSO,
	long = least absolute shrinkage and selection operator}	

\DeclareAcronym{LR}{short = LR,
	long = logistic regression}
		
\DeclareAcronym{PCA}{short = PCA,
	long = principal component analysis}
	
\DeclareAcronym{PLS}{short = PLS,
	long = partial least squares}	
	
\DeclareAcronym{RBC}{short = RBC,
	long = reconstruction-based contribution}

\DeclareAcronym{ANFIS}{short = ANFIS,
	long = adaptive network fuzzy inference system}
	
\DeclareAcronym{ANN}{short = ANN,
	long = artificial neural network,
	short-indefinite = an}	

\DeclareAcronym{BN}{short = BN,
	long = Bayesian network}	
		
\DeclareAcronym{CNN}{short = CNN,
	long = convolutional neural network}	
	
\DeclareAcronym{DNNE}{short = DNNE,
	long = decorrelated neural network ensemble}		
	
\DeclareAcronym{DNN}{short = DNN,
	long = deep neural network}	

\DeclareAcronym{ELM}{short = ELM,
	long = extreme learning machine}
			
\DeclareAcronym{GAN}{short = GAN,
	long = generative adversarial network}

\DeclareAcronym{GPR}{short = GPR,
	long = Gaussian process regression}	
	
\DeclareAcronym{GRNN}{short = GRNN,
	long = general regression neural network}	
	
\DeclareAcronym{MLP}{short = MLP,
	long = multilayer perceptron,
	short-indefinite = an}

\DeclareAcronym{RBFNN}{short = RBFNN,
	long = radial basis function neural network,
	short-indefinite = an}
		
\DeclareAcronym{RNN}{short = RNN,
	long = recurrent neural network,
	short-indefinite = an}
	
\DeclareAcronym{RT}{short = RT,
	long = regression tree,
	short-indefinite = an}			

\DeclareAcronym{RVM}{short = RVM,
	long = relevance vector machine,
	short-indefinite = an}		
	
\DeclareAcronym{SFA}{short = SFA,
	long = slow feature analysis}	

\DeclareAcronym{SVM}{short = SVM,
	long = support vector machine}
	
\DeclareAcronym{TL}{short = TL,
	long = transfer learning}	
		
\DeclareAcronym{VAE}{short = VAE,
	long = variational autoencoder}	
	
\DeclareAcronym{WNN}{short = WNN,
	long = wavelet neural network}

\DeclareAcronym{A3C}{short = A3C,
	long = asynchronous advantage actor-critic}	

\DeclareAcronym{ADP}{short = ADP,
	long = approximate dynamic programming}

\DeclareAcronym{DDPG}{short = DDPG,
	long = deep deterministic policy gradient}
		
\DeclareAcronym{DQN}{short = DQN,
	long = deep $Q$-network}

\DeclareAcronym{HJB}{short = HJB,
	long = {Hamilton-Jacobi-Bellman}}
	
\DeclareAcronym{MPC}{short = MPC, 
	long = model predictive control,
	long-plural-form = model predictive controllers,
	short-indefinite = an}		

\DeclareAcronym{PI2}{short = $\text{PI}^2$,
	long =policy improvement with path integrals}
	
\DeclareAcronym{PID}{short = PID, 
	long = proportional-integral-derivative}
	
\DeclareAcronym{PI}{short = PI, 
	long = proportional-integral}
			
\DeclareAcronym{PPO}{short = PPO,
	long = proximal policy optimization}

\DeclareAcronym{REINFORCE}{short = REINFORCE,
	long = {\emph{RE}ward Increment $=$ \emph{N}onnegative \emph{F}actor $\times$ \emph{O}ffset \emph{R}einforcement $\times$ \emph{C}haracteristic \emph{E}ligibility}}
	
\DeclareAcronym{RTO}{short = RTO,
	long = real-time optimization,
	short-indefinite = an}	
	
\DeclareAcronym{SAC}{short = SAC,
	long = soft actor-critic}	
		
\DeclareAcronym{TD3}{short = TD3,
	long = twin-delayed DDPG}

\DeclareAcronym{GP}{short = GP,
	long = Gaussian process,
	long-plural-form = Gaussian processes}
	
\DeclareAcronym{RBF}{short = RBF,
	long = radial basis function,
	short-indefinite = an}	
	
\DeclareAcronym{SAE}{short = SAE,
	long = sparse autoencoder}	
		
\DeclareAcronym{DBN}{short = DBN,
	long = deep belief network}	
	
\DeclareAcronym{LSTM}{short = LSTM,
	long = long short-term memory}

\DeclareAcronym{KL}{short = KL,
	long = Kullback-Leibler}
	
\DeclareAcronym{MDP}{short = MDP,
	long = Markov decision process,
	long-plural-form = Markov decision processes,
	short-indefinite = an}
	
\DeclareAcronym{LQR}{short = LQR, 
	long = linear quadratic regulator}
	
\DeclareAcronym{LTI}{short = LTI, 
	long = linear time-invariant,
	short-indefinite = an}
	
\DeclareAcronym{GPS}{short = GPS,
	long = guided policy search}	
	
\DeclareAcronym{GRU}{short = GRU,
	long = gated recurrent unit}	
	
\DeclareAcronym{ESN}{short = ESN,
	long = echo state network}	

\DeclareAcronym{ENN}{short = ENN,
	long = Elman neural network}	
\journal{arXiv}

\begin{document}

\begin{frontmatter}



%\title{Modern machine learning techniques for monitoring and control of industrial processes: A survey}
%\title{\texorpdfstring{{\color{red}A practical survey of machine learning techniques for industrial monitoring and control}}}
\title{Machine learning for industrial sensing and control: A survey and practical perspective\tnoteref{label1}}
\tnotetext[label1]{Please cite the journal version in Control Engineering Practice: \url{https://doi.org/10.1016/j.conengprac.2024.105841}}

% \author{}
% \thanks[footnoteinfo]{Email:bhushan.gopaluni@ubc.ca}
\author[UBC-math]{Nathan P. Lawrence}
\ead{input@nplawrence.com}
\author[UA]{Seshu Kumar Damarla}
\author[Ninth]{Jong Woo Kim}
\author[UA]{Aditya Tulsyan} % TODO update
\author[UA]{Faraz Amjad}
\author[Eighth]{Kai Wang}
\author[Imperial]{Benoit Chachuat}
\author[SNU]{Jong Min Lee}
\author[UA]{Biao Huang}
\author[UBC-chbe]{R. Bhushan Gopaluni}
\ead{bhushan.gopaluni@ubc.ca}
% \address{}

\address[UBC-math]{Department of Mathematics, University of British Columbia, Canada}
\address[UBC-chbe]{Department of Chemical and Biological Engineering, University of British Columbia, Canada}
\address[UA]{Department of Chemical and Materials Engineering, University of Alberta, Canada}
\address[Imperial]{The Sargent Centre for Process Systems Engineering, Department of Chemical Engineering, Imperial College London, London SW7 2AZ, UK}
\address[SNU]{School of Chemical and Biological Engineering, Institute of Chemical Processes, Seoul National University, 1, Gwanak-ro, Gwanak-gu, Seoul 08826, Republic of Korea}
\address[Ninth]{Department of Energy and Chemical Engineering, Incheon National University, Incheon 22012, Republic of Korea}
\address[Eighth]{School of Automation, Central South University, Changsha, 410083, China}

\begin{abstract}

With the rise of deep learning, there has been renewed interest within the process industries to utilize data on large-scale nonlinear sensing and control problems.
We identify key statistical and machine learning techniques that have seen practical success in the process industries.
To do so, we start with hybrid modeling to provide a methodological framework underlying core application areas: soft sensing, process optimization, and control.
Soft sensing contains a wealth of industrial applications of statistical and machine learning methods.
We quantitatively identify research trends, allowing insight into the most successful techniques in practice.
 We consider two distinct flavors for data-driven optimization and control: hybrid modeling in conjunction with mathematical programming techniques and reinforcement learning.
Throughout these application areas, we discuss their respective industrial requirements and challenges.
 A common challenge is the interpretability and efficiency of purely data-driven methods. 
This suggests a need to carefully balance deep learning techniques with domain knowledge.
As a result, we highlight ways prior knowledge may be integrated into industrial machine learning applications.
The treatment of methods, problems, and applications presented here is poised to inform and inspire practitioners and researchers to develop impactful data-driven sensing, optimization, and control solutions in the process industries.
\end{abstract}


\end{frontmatter}

%\linenumbers

% \section{}
% \label{}

\section{Motivation}
\label{sec:motivation}
\acbarrier

Data analytics and \ac{ML} ideas are not new to the process industries\footnote{For simplicity, we refer to chemical and biological industries as \emph{process industries} or just \emph{industries}.}. The review paper by \citet{venkatasubramanian2019promise} provides an excellent overview of the history, successes, and failures of various attempts over more than three decades to use ideas from \ac{AI} in the industry. In particular, statistical techniques such as \ac{PCA}, \ac{PLS}, \ac{CCA}, and time series methods for modeling, such as maximum likelihood estimation and prediction error methods, have been extensively used in industry \citep{chiang2000fault}. Several classification and clustering algorithms, such as $k$-means, \acp{SVM}, and Fisher discriminant analysis, are also widely used in industry \citep{qin2019advances,ge2017DataMining}. And several nonlinear approaches, such as kernel methods, \acp{GP}, and adaptive control algorithms, such as reinforcement learning, have been applied in some niche applications \citep{shin2019reinforcement,spielberg2019SelfDriving, tulsyan2019automatic}.\par  
%Most of these algorithms do not scale well with data. \par

Despite the longstanding success of many statistical techniques in industry, there is also considerable interest in developing sensing and control technologies based on more recent \ac{ML} architectures \citep{venkatasubramanian2019promise,nian2020review,bi2022one}.
Broadly speaking, these aspirations are driven by the promises of increased autonomy: increased operational efficiency, consistency, and safety; improved scalability beyond linear methods; upskilling of plant personnel \citep{gamer2020autonomous}.
Consequently, this paper addresses the need to dissect and organize the general use of modern \ac{ML} techniques in industrial applications.
In doing so, such a treatment will inform practitioners of the latest research trends and their potential practical impact. 
Conversely, researchers in core areas will benefit from a holistic view of successful \ac{ML} techniques and the industrial requirements they satisfy.

%The rest of the article is divided as follows: \cref{sec:softSensor} surveys statistical and deep learning methods for soft sensor development in process industries; \cref{sec:faultDetection} discusses deep learning methods for process modeling and process monitoring; \cref{sec:RL} surveys reinforcement learning methods for process control applications; \cref{sec:hybrid} discusses hybrid modeling techniques for process optimization and control applications; and \cref{sec:smallData} surveys applications of ML tools in applications with limited (or small) datasets. 
%This is primarily a problem-driven survey; however, we have provided sufficient references for interested readers on the underlying methods discussed here. Moreover, we have included additional exposition on some of these methods in the supplementary material.

\subsection{Overview and scope}

This paper is a significant extension of \citet{gopaluni2020modern}: in addition to a more detailed and expansive treatment of the literature, we discuss the practical success of various methods.  
Note that this is primarily a problem-driven survey; however, we have provided sufficient references for interested readers on the underlying methods discussed here. 
Moreover, we have included additional exposition on some of these methods in the supplementary material.

Hybrid modeling is first introduced to provide a conceptual framework underlying core application areas, namely:\footnote{Tackling the trio of sensing, monitoring, and control in detail is beyond the scope of this paper. Readers are referred to \citet{ge2017DataMining,bi2022one,sansana2021RecentTrends} for more details on \ac{ML} methods for process monitoring.}
\begin{enumerate}
	\item Soft sensing
	\item Process control
\end{enumerate}
Process control also includes process optimization.
In our survey, we identify several methodological areas of research: statistical learning and machine learning, deep learning and its variants, and reinforcement learning.
Algorithms from each of these methodological areas are used to varying degrees among the core applications.
Soft sensing encompasses more statistical and machine learning methods, with some discussion of deep learning.
%While process monitoring utilizes soft sensing techniques, it also focuses more on deep learning methods.
On the optimization and control side, we discuss hybrid modeling in tandem with mathematical programming and reinforcement learning.

This is by no means an exhaustive survey of the recent research on these topics. 
However, we have tried our best to include some of the most critical developments of \ac{ML} tools in the process industries.
In that vein, we only discuss methods that have seen industrial use or have received considerable research attention within process systems engineering, either in real life or in simulations.
Therefore, speculation about the potential use of very recent developments in the broader \ac{ML} community, such as ChatGPT or other large language models, is beyond the scope of this paper.
However, we provide insight into the practical deployment of \ac{ML} techniques in the process industries.
\par

This paper surveys a large number of algorithms.  
\Cref{fig:biao5} gives a convenient list to reference across all sections.
Throughout this paper, \emph{\acl{AI}} is the broadest term for classifying machines that aim to mimic human intelligence. It is intended to predict, automate, and optimize the tasks humans have traditionally performed, such as speech recognition, image recognition, decision-making, and translation.
\emph{\Acl{ML}} is an area of artificial intelligence and computer science where algorithms are developed to extract patterns from data and make predictions.
\emph{Supervised learning} is a branch of \acl{ML} comprised of algorithms for determining a predictive model based on labeled data with known outcomes.
On the other hand, \emph{unsupervised learning} is a branch of \acl{ML} devoted to learning patterns from unlabeled data.

\begin{table*}[tbp]
\caption{Full forms for acronyms. Divided into three sections, top to bottom: 1) statistical learning, 2) machine learning \& deep learning, and 3) reinforcement learning \& control methods.}
\begin{center}
{\footnotesize
    \begin{tabularx}{\textwidth}{l L l L}
      \toprule % <-- Toprule here
      \acs{CCA} & \Acl{CCA} & \acs{LASSO} & \Acl{LASSO} \\
      \acs{FA} & \Acl{FA} & \acs{LR} & \Acl{LR} \\
      \acs{GMM} & \Acl{GMM} & \acs{PCA} & \Acl{PCA} \\
      \acs{ICA} & \Acl{ICA} & \acs{PLS} & \Acl{PLS} \\
      \acs{LARS} & \Acl{LARS} & \acs{RBC} & \Acl{RBC} \\
      \midrule
      \acs{ANFIS} & \Acl{ANFIS} & \acs{GRNN} & \Acl{GRNN} \\
       \acs{ANN} & \Acl{ANN}  & \acs{MLP} & \Acl{MLP} \\
    \acs{BN} & \Acl{BN}  & \acs{RBFNN} & \Acl{RBFNN} \\
      \acs{CNN} & \Acl{CNN} & \acs{RNN} & \Acl{RNN} \\
      \acs{DNNE} & \Acl{DNNE} & \acs{RT} & \Acl{RT} \\
      \acs{DNN} & \Acl{DNN}  & \acs{RVM} & \Acl{RVM} \\
      \acs{ELM} & \Acl{ELM} & \acs{SFA} & \Acl{SFA} \\
      \acs{ESN} & \Acl{ESN} & \acs{SVM} & \Acl{SVM} \\
     \acs{ENN} & \Acl{ENN} & \acs{TL} & \Acl{TL} \\
       \acs{GPR} & \Acl{GPR}    & \acs{WNN} & \Acl{WNN} \\
      \midrule
      \acs{A3C} & \Acl{A3C} & \acs{PI2} & \Acl{PI2} \\
      \acs{ADP} & \Acl{ADP} & \acs{PID} & \Acl{PID} \\
      \acs{DDPG} & \Acl{DDPG} & \acs{PPO} & \Acl{PPO} \\
      \acs{DQN} & \Acl{DQN} & \acs{RTO} & \Acl{RTO} \\
      \acs{HJB} & \Acl{HJB} & \acs{SAC} & \Acl{SAC} \\
      \acs{MPC} & \Acl{MPC} & \acs{TD3} & \Acl{TD3} \\
      \bottomrule % <-- Bottomrule here
    \end{tabularx}
    }
    \end{center}
    \label{fig:biao5}  
\end{table*}

\section{Mathematical modeling approaches}
\label{sec:hybrid}
\acbarrier

The core applications of this paper are soft sensing and process optimization and control.
These areas rely on dynamic mathematical models to infer measurements, make decisions, and synthesize controllers.
Therefore, before describing the prominent \ac{ML} techniques in these areas, it is useful to introduce the foundational assumptions and architectures underlying such models.

\subsection{Knowledge-driven, data-driven, and hybrid modeling}

{\em Knowledge-driven} (mechanistic or white box) modeling based on first principles and {\em data-driven} (or black box) modeling constitute two opposite strategies. Developing mechanistic models requires a deep understanding of the processes at play. It is often labor-intensive, but embodying first principles may enable extrapolation beyond the conditions under which these models are trained. By construction, mechanistic models have a fixed structure and comprise a fixed number of parameters, often with a physical or empirical interpretation. For this reason, they may also be classified as {\em parametric} models.

By contrast, data-driven models require little physical knowledge and are fast to deploy or maintain. But a larger dataset is also typically needed for their construction, and their validity may not extend far beyond the conditions under which they are trained. The structure of a data-driven model does not need to be dictated by \emph{a priori} knowledge but may be tailored to the training data at hand instead. A further distinction is whether a data-driven model tries to describe data with a set of parameters of fixed size, regardless of the size of the training dataset, in which case it is categorized as {\em parametric}, or whether its structure and number of parameters may evolve with the size of the dataset, commonly referred to as {\em nonparametric} \citep{hastie2009elements,goodfellow2016deep}. The so-called nonparametric regression models fall in the second category, whereby the predictor does not take a predetermined form, using techniques such as nearest-neighbor interpolation, local regression, and \ac{GP} regression. However, the distinction between parametric and nonparametric models in statistical and \ac{ML} is not without intricacies. For instance, a linear \acs{SVM} is a typical example of a parametric model, having a fixed number of parameters---a weight for each input dimension. In contrast, \acs{RBF}-kernel \acs{SVM} may be considered nonparametric since the number of parameters grows with the size of the training set---a weight for each training point.

The basic idea behind {\em hybrid} models is to combine knowledge-driven and data-driven models in such a way as to overcome their respective limitations. This strategy is also frequently referred to as {\em gray box} or {\em block-oriented} modeling in the literature. At the same time, the term hybrid {\em semi-parametric} modeling is coined to describe those hybrid models where the data-driven component is nonparametric \citep{VonStosch2014}. {\em Multi-fidelity modeling} has also developed fast in recent years and is akin to hybrid modeling. The idea is to use a (possibly inaccurate) knowledge-driven model as low-fidelity and correct it with (noisy) process data, considered to be higher fidelity \citep{Peherstorfer2018}. In particular, this strategy has been applied in uncertainty propagation, inference, and optimization and is also instrumental in small data problems (see supplementary material).

It is worth noting that hybrid modeling has been investigated for over $25$ years in chemical and biological process engineering \citep{Psichogios1992,Su1992,Thompson1994,Chen2000}. The claimed benefits of hybrid modeling in these application domains include faster prediction capability, better extrapolation capability, better calibration properties, easier model life-cycle management, and higher benefit/cost ratio to solve complex problems; see recent survey papers on the development and applications of hybrid models by \citet{VonStosch2014,Solle2017,Schuppert2018,Zendehboudi2018,Ahmad2020,bradley2022}. Hybrid models may be used to enable soft sensors (see \cref{sec:softSensor}) or model-based optimization and control (see \cref{sec:optim&control}) in a first principles approach. 

\begin{figure}[tbp]
    \centering
    \includegraphics[width=8.4cm]{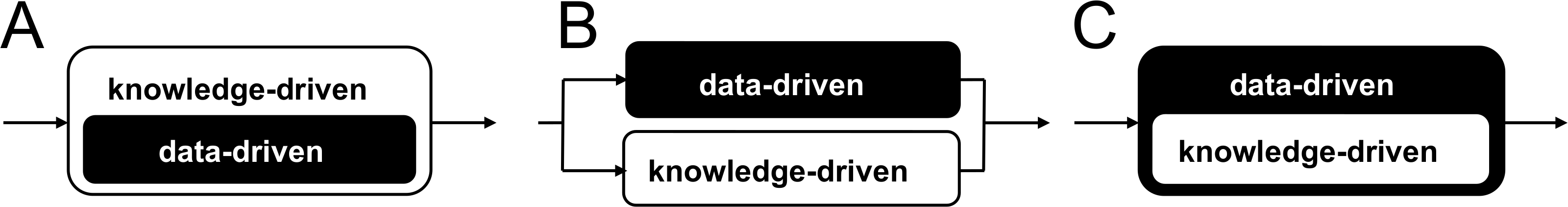}
    \caption{Typology of hybrid models (see \citet{VonStosch2014}). \textsf{A} and \textsf{C} represent serial structures: under \textsf{A}, a data-driven model is used as input to a knowledge-driven model; \textsf{C} is the reverse. \textsf{B} represents a parallel structure in which knowledge-driven predictions are corrected by data-driven predictions.}
    \label{fig:chachuat1}
\end{figure}

\subsection{Hybrid modeling paradigms}

\subsubsection{Traditional serial and parallel hybrid models}

The usual classification of hybrid model structures is either as {\em serial} or {\em parallel} \citep{Agarwal1997}. In the serial approach, the data-driven model is most commonly used as an input to the mechanistic model (see \cref{fig:chachuat1}\textsf{A}), for instance, a material balance equation with a kinetic rate expressed using a data-driven model. This structure is especially suited to situations where precise knowledge about specific underlying mechanisms is lacking, yet sufficient process data exists to infer the corresponding relationship \citep{Psichogios1992,Chen2000}. However, when the mechanistic part of the model presents a structural mismatch, one should not expect the serial approach to perform better than a purely mechanistic approach. In the parallel approach, by contrast, the output of the data-driven model is used to correct the predictions of the mechanistic model \citep{Su1992,Thompson1994}, most often in the form of an additive correction (see \cref{fig:chachuat1}\textsf{B}). This structure can significantly improve the prediction accuracy of a mechanistic model when the data-driven component is trained on the residuals between process observations and mechanistic model predictions. However, this accuracy may not be better than the sole mechanistic model when the process conditions differ drastically from those in the training set. 

Historically, the most common data-driven modeling techniques embedded in hybrid models have been \ac{MLP} and \ac{RBF}-based regression \citep{VonStosch2014}. Recent representative applications include the development of a serial hybrid model to predict hydraulic fractures created by injecting fluid into a reservoir that accounts for the leak-off rate of the fracturing fluid using \iac{MLP} \citep{Bangi2020} and the development of a serial hybrid model of the thin film growth process coupling a macroscopic gas phase model described by partial differential equations to a microscopic thin-film model described by stochastic partial differential equations via \iac{MLP} \citep{Chaffart2018}. Naturally, many other statistical and \ac{ML} techniques have also been investigated in this context. For instance, \citet{Ghosh2019} used subspace identification to construct the data-driven component in a parallel hybrid model and demonstrated the approach on a batch polymerization reactor. \citet{Lopez2020} developed a serial hybrid model of a lignocellulosic fermentation process, whereby the glucose concentration is estimated from spectroscopic data using a \ac{PLS} regression model. \Ac{GP} regression has also attracted attention due to its ability to estimate the predictor's variance, for example, in bioprocess engineering applications \citep{Zhang2019b}. 

Parallel hybrid models can significantly alleviate the issue of maintaining a complex mechanistic model since the data-driven component is trained to capture model mismatch in the first place, possibly in a nonparametric manner. For dynamic systems in particular, a popular approach entails training the data-driven model on the residuals between the predicted and observed states at given time instants \citep{Duarte2004}. Notice that such a data-driven model could either comprise algebraic or differential equations. By contrast, serial hybrid models can prove more challenging to design, especially when the outputs of the data-driven component cannot be observed directly \citep{dePrada2018}. In such a case, training and assessing the performance of the data-driven component requires one to simulate the full serial hybrid model and compare its outputs to the available observations. Identifying the unknown model parameters within such hybrid models has relied on regularized regression techniques, such as \acs{LASSO} and \acs{LARS} \citep{Hesterberg2008}. 
% See also \citet{AlMatouq2020} for a recent application of this approach to identifying homogeneous reaction systems.

Another challenge shared by serial and parallel hybrid modeling paradigms is automatically detecting the best structure for the data-driven component. Generally speaking, minimizing the number of parameters needed to capture the underlying mechanisms is desirable, that is, to neither underfit nor overfit the data. Classical approaches to help discriminate among multiple nonparametric model structures include the Akaike Information Criterion and Bayesian Information Criteria. \citet{Willis2017} proposed an approach based on sparse regression and mixed-integer programming to simultaneously decide the structure and identify the parameters for a class of rational functions embedded into a serial hybrid model. Recently, \citet{Zhang2020} applied hybrid modeling in combination with sparse identification of nonlinear dynamics \citep[SINDy;][]{Brunton2016} to a photo-production bioprocess, whereby a sparse quadratic correction of the kinetic model is identified using mixed-integer nonlinear programming techniques. More generally, there is significant scope for extending sparse and symbolic regression techniques to enable the construction of hybrid models. Notably, the platform ALAMO \citep{Wilson2017} can enforce constraints on the response variables to incorporate first principles knowledge, thereby revealing hidden relationships between regression parameters that may not be directly available to the modeler. One approach to incorporating such constraints is via semi-infinite programming \citep{Cozad2015}. 
% which can prove computationally challenging in practice. 
Another promising direction entails using sum-of-squares optimization techniques to tackle this problem~\citep{Nauta2007,Pitarch2019}.

\subsubsection{Emerging trends}

The traditional hybrid modeling approach has put a mechanistic model at its core. It uses data-driven elements to either describe specific unknown or poorly understood mechanisms or correct the predictions of the mechanistic model. Another way of incorporating domain knowledge and mechanistic models is feature engineering, where the inputs to the data-driven elements are augmented by terms that would also appear in mechanistic models; for instance, think of enthalpy, which is not a measurement but a useful term in energy balances. Hybrid models whereby the mechanistic model is now used as an input to the data-driven component have become increasingly popular in recent years (see \cref{fig:chachuat1}\textsf{C}). This approach includes physics-informed neural networks where the underlying conservation equations are imposed as extra constraints on the \ac{MLP}'s parameters \citep{Raissi2019}, like the classical orthogonal collocation theory on finite elements using piecewise polynomials \citep{Carey1975,biegler2007overview}. Co-Kriging techniques have also been developed where a \ac{GP} trained using data from a mechanistic model is combined with a second \ac{GP} trained using process data (or a high-fidelity model) \citep{Liu2018}. Such an approach also enables multi-fidelity modeling using linear or nonlinear autoregressive techniques \citep{LeGratiet2014,Perdikaris2017} and deep \acp{GP} \citep{Cutajar2018}, and finding applications, for instance, in the optimization of complex black box simulators and legacy codes. Another body of research has been concerned with learning a dynamic system by accounting for prior information, for instance, the regression of polynomial dynamic systems with prior information using sum-of-squares optimization methods \citep{Ahmadi2020}.

Since there is no universal framework, a recurring challenge with hybrid modeling is selecting the appropriate paradigm---for example, physics-driven against data-driven backbone, or serial against parallel structure---for a particular application, such as small vs. large datasets or noisy vs. high-quality data. This selection process still lacks a solid theoretical basis, although systematic computational comparisons of various hybridization techniques have emerged in recent years \citep{bradley2022}. Finally, looking beyond current hybrid models, \citet{venkatasubramanian2019promise} argued for the development of hybrid \ac{AI} systems that would combine not only mechanistic with data-driven models but also causal models-based explanatory systems or domain-specific knowledge engines. Likewise, the mechanistic model could be replaced by a graph-theoretical model, such as signed digraphs, or a production system model, creating entirely new research fields.

\section{Soft sensors in process industries}
\label{sec:softSensor}
\acbarrier

Soft sensing represents the most fundamental application of \ac{ML} techniques in the process industries.
By extension, optimization and control add complexity to a soft sensing core.
As a result, based on our analysis and own experience, soft sensing contains the most industrial penetration of \ac{ML} applications.
We quantitatively analyze which \ac{ML} methods have seen practical success and which are currently being researched.
We offer practical considerations and insights for implementing soft sensors in practice to balance the apparent industrial-academic disconnect.

\subsection{Motivation for soft sensing}

\begin{figure}[tbp]
    \centering
    \includegraphics[width=8.4cm]{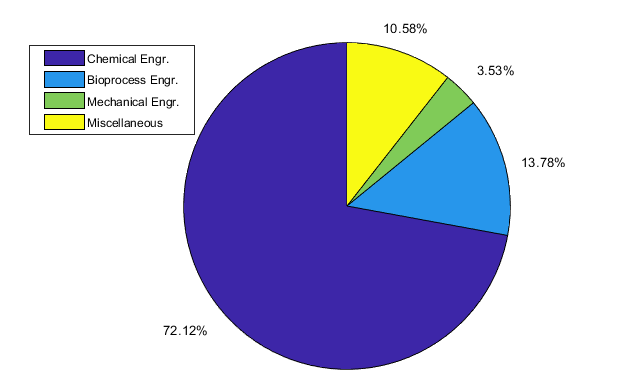}
    \caption{Distribution of soft sensor applications.}
    \label{fig:biao7}
\end{figure}

In the process industries, some variables are difficult to measure online due to technological limitations or the high cost of sensors. These variables indicate a product's intermediate or final quality and must be continuously monitored and controlled. In such circumstances, mathematical models are developed using easy-to-measure variables. These models provide a continuous estimate for quality variables in real time. The mathematical models devoted to the estimation of plant variables are called \emph{soft sensors} \citep{khatibisepehr2013design, fortuna2007soft}.
The process industries, such as refineries, steel plants, polymer industries, or cement industries, remain the dominant users of soft sensors (see \cref{fig:biao7}). 

Similar to hybrid modeling, soft sensors can be categorized as \emph{knowledge-driven} and \emph{data-driven}. Knowledge-driven soft sensors (or white box models), such as Kalman filters, are based on first principles models that describe the physical and chemical laws that govern the process, such as mass and energy balance equations. In contrast, data-driven soft sensors (or black box models) have no information about the process and are based on empirical observations (historical process data). A third type of soft sensor, called \emph{hybrid models} (or gray box models), uses a data-driven method to estimate the parameters of a knowledge-driven model. This special combination is closely related to the general concept of \emph{hybrid modeling}, as discussed in \cref{sec:hybrid}. For instance, a model may incorporate physics-based simulations and process measurements.

\par

\subsection{A quantitative overview of soft sensing}

\begin{figure}[tbp]
    \centering
    \includegraphics[width=8.4cm]{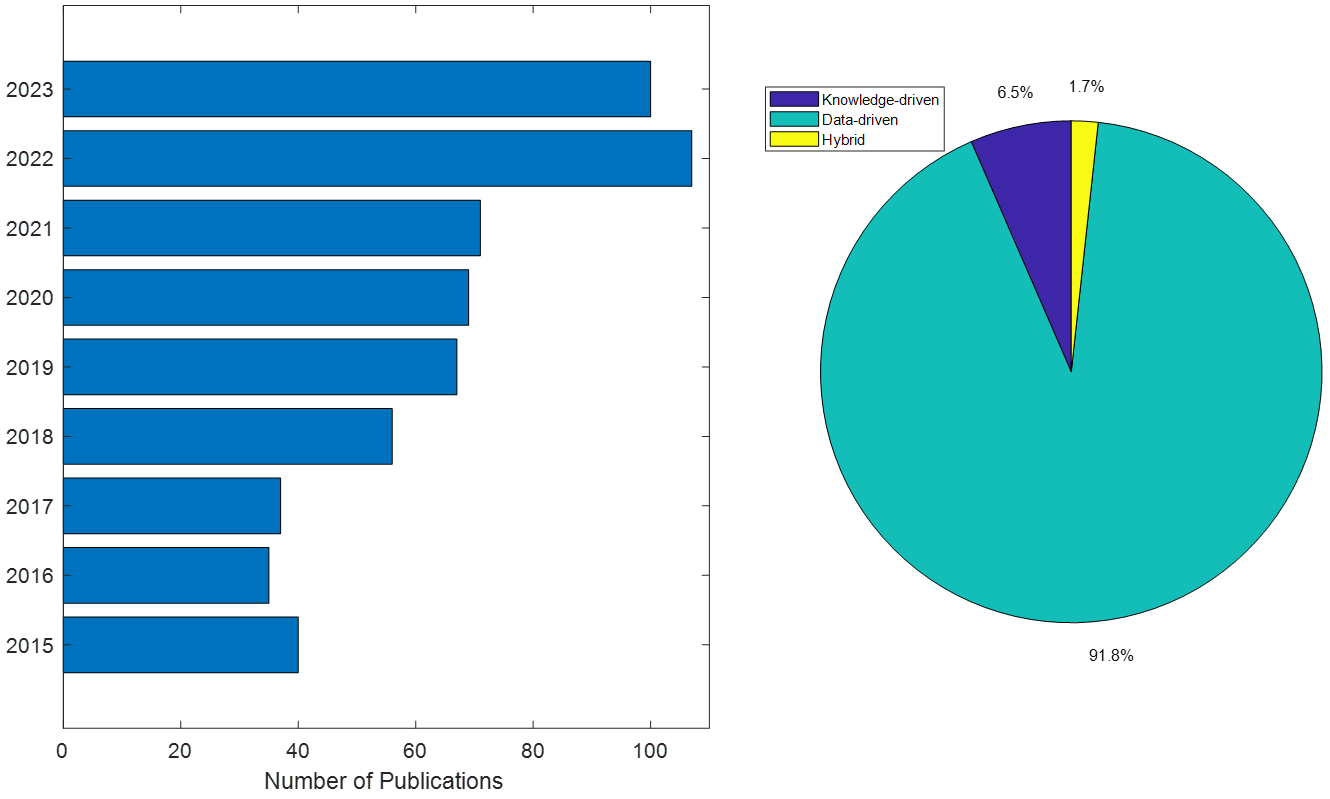}
    \caption{Research publication in soft sensors from $2015$ to $2023$.}
    \label{fig:biao1}
\end{figure}

Literature was collected by gathering articles published between $2015$ and $2023$ in relevant journals from publishing houses like Elsevier, Springer, Wiley, Taylor and Francis, MDPI, World Scientific, Hindawi, De Gruyter, AMSE, and IEEE. For the publication search, keywords such as ``soft sensor'', ``virtual sensor'' or ``inferential model'' were used. 
The statistics shown in \cref{fig:biao1} were computed based on the collected literature.

These statistics indicate that the research conducted in soft sensing between $2015$ and $2023$ was primarily focused on data-driven models. This is unsurprising, as data-driven soft sensors can often capture complex and unexplained process dynamics more succinctly. In contrast, knowledge-driven soft sensors require much expert process knowledge, which is not always available. In addition, knowledge-driven soft sensors are difficult to calibrate, especially for complex nonlinear processes. Note that hybrid model-based soft sensors received the least research attention. Data-driven soft sensors can be further categorized based on the learning technique used for modeling.

%\Cref{fig:biao2} and 
\Crefrange{table:dist_SL}{table:dist_ANN} show the current trends in the data-driven soft sensing.
\Cref{fig:biao5} contains the full forms for the acronyms used in \crefrange{table:dist_SL}{table:dist_ANN}.
The research in soft sensing has dramatically shifted from statistical to \ac{ML} methods.
\Acp{ANN} received the greatest attention among \ac{ML} methods. 
The class of feedforward single hidden layer neural networks (shallow networks)---encompassing \ac{MLP}, \acs{GRNN}, \acs{ELM}, \ac{RBFNN}, \ac{WNN} in \cref{fig:biao5}---have more applications in soft sensing than \acp{RNN} and deep learning. 
Aside from \acp{ANN}, \ac{SVM} is the second most widely used \ac{ML} method for developing inferential models.\par

\begin{table*}[tbp]
\caption{Distribution of data-driven methods for soft sensors, split between statistical and \acs{ML} methods.}
\begin{center}
\begin{tabular}{ll | ll}
\toprule
Statistical methods & \% of publications & \Acs{ML} methods & \% of publications \\
\midrule
PLS      & 11.38 & ANN    & 47.72                 \\
PCA      & \phantom{0}4.54 & TL & \phantom{0}2.02 \\
FA       & \phantom{0}0.95 & RT     & \phantom{0}4.59  \\
ICA      & \phantom{0}1.70 & SFA    & \phantom{0}2.75 \\
LASSO    & \phantom{0}1.51 & RVM    & \phantom{0}2.57 \\
GMM      & \phantom{0}2.64 & SVM    & \phantom{0}7.53  \\
- & \phantom{0}- & ANFIS  & \phantom{0}1.65   \\
- & \phantom{0}- & GPR    & \phantom{0}5.87 \\
- & \phantom{0}- & BN     & \phantom{0}2.57 \\
\midrule
Total    & 22.73 & Total & 77.27 \\                 
\bottomrule
\end{tabular}
\end{center}
\label{table:dist_SL}  
\end{table*}

\begin{table}[tbp]
\caption{Distribution of various types of \acsp{ANN} for soft sensors.}
\begin{center}
\begin{tabular}{ll}
\toprule
Method & \% of publications \\
\midrule
MLP    &14.23                 \\
DNNE     & \phantom{0}0.75                \\
DNN     & 34.64                  \\
ELM    & 11.92                  \\
GRNN    & \phantom{0}6.92                  \\
WNN    & \phantom{0}2.31                  \\
RBFNN  & \phantom{0}2.68                  \\
RNN    & 21.94                 \\
ENN    & \phantom{0}0.38                   \\
ESN    & \phantom{0}4.23                   \\
%\midrule
%Total  & 100.00   \\              
\bottomrule
\end{tabular}
\end{center}
\label{table:dist_ANN}
\end{table}

\begin{figure}[tbp]
    \centering
    \includegraphics[width=8.4cm]{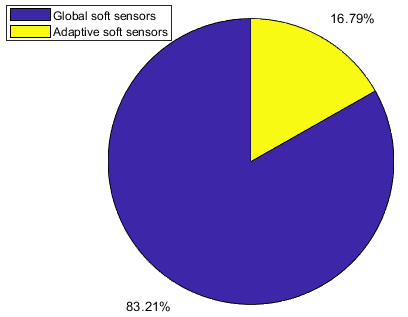}
    \caption{Distribution of global and adaptive soft sensors.}
    \label{fig:biao6}
\end{figure}

\Acl{TL} is slowly gaining applications in inferential measurements.
\Acl{TL} alludes to the scenario where knowledge gained while performing one specific task is exploited to carry out a different but related task. 
Especially when data collection becomes difficult in the task of interest, \acl{TL} still works by sharing information on relevant data in other domains \citep{chu2019transfer}. 
%However, \acl{TL} is more suitable for predictive modeling problems that use image data as input. 
\Acl{TL} has yet to be applied to the online prediction of process variables.

Static (time-invariant) soft sensors are developed using data from a single operating mode.
However, their prediction accuracy degrades over time as the process shifts to a new operating region.
Adaptive soft sensors tackle this issue by updating their parameters based on new samples.\footnote{One may further distinguish between unimodal and multimodal soft sensors. For instance, a static soft sensor may be developed for multimodal distributions. In any case, we are referring to the case of a static soft sensor degrading over time.}
Less than one-third of soft sensors are adaptive, most of which use a just-in-time strategy to update model parameters in response to samples arriving in real time (see \cref{fig:biao6} and \cref{table:dist_locmod}). 
Therefore, computationally feasible methods are required. 
In particular, \ac{PLS} is the preferred algorithm for local modeling. 
%Over two-thirds of the publications reported new theoretical developments for better inferential measurements. 
The training of global soft sensor models is performed offline. 
Then, trained soft sensor models are deployed online to obtain real-time estimates for key process or quality variables. 
Although training time is comparatively very high, most global soft sensors produce estimates quickly when used online \citep{fortuna2007soft,kadlec2009data} 

In \cref{table:biao2}, publications on each data-driven technique have been grouped into three categories: publications based on simulation data, publications based on industrial data, and publications that reported industrial implementation. Notice that most soft sensors have been developed and tested on industrial data. Still, only some of them---\ac{PLS}, \ac{MLP}, \ac{WNN}, \ac{SVM}, \ac{RVM}, \ac{GPR} and \ac{RT}---have made it into actual industrial implementation.
Of course, there may be a publication bias for academic examples, as not all real-world industrial applications may be reported on.

\begin{table}[tbp]
\caption{Distribution of statistical and \acs{ML} methods in local modeling of adaptive soft sensors.}
\begin{center}
\begin{tabular}{ll}
\toprule
Method & \% of publications \\
\midrule
PCA    & \phantom{0}7.05                 \\
MLP     & \phantom{0}3.52                  \\
SVR     & 11.77                  \\
GMM    & \phantom{0}2.35                 \\
GPR    & 15.30                  \\
BN    & \phantom{0}4.70                 \\
RVM  & \phantom{0}2.35                \\
ELM    & \phantom{0}9.42                  \\
FA    & \phantom{0}3.52 \\
PLS   & 31.78 \\
LASSO & \phantom{0}8.24 \\           
\bottomrule
\end{tabular}
\end{center}
\label{table:dist_locmod}  
\end{table}

\begin{table}[tbp]
\caption{Breakdown of methods for soft sensors according to the level of industrial applications.}
\begin{center}
\begin{tabularx}{\linewidth}{XRRR|R} 
\toprule
 Method & {Simulation data} & {Industrial data} & {Industrial use} & {Number of publications}\\ 
 \midrule
PCA	& 8 	& 16	&0	& 24\\
PLS	&12	&45	&3	&60\\
SFA	&4	&11	&0	&15\\
ICA	&7	&2	&0	&9\\
LR	&0	&1	&0	&1\\
LASSO	&1	&6&0	&7\\
FA	&0	&5	&0	&5\\
GMM	&3	&11	&0	&14\\
MLP	&4	&32	&1	&37\\
RBFNN	&2	&5	&0	&7\\
WNN	&1	&4	&1	&6\\
RNN	&4	&53	&0	&57\\
GRNN	&5	&13	&0	&18\\
ELM	&6	&25	&0	&31\\
ANFIS	&3	&6&0	&9\\
DNNE &1	&1	&0	&2\\
DNN	&7	&80	&0	&87\\
SVM	&12	&21	&4&37\\
RVM	&4	&9&1&14\\
GPR	&12	&18	&2	&32\\
RT	&5	&19	&1	&25\\
BN	&0	&14	&0	&14\\
TL	&4	&7	&0	&11\\
ESN &2  &9  &0 &11\\
ENN  &0  &1  &0 &1\\
\midrule
Total    & 107 & 414 & 13 & 534 \\                 
\bottomrule
\end{tabularx}
\label{table:biao2}
\end{center}
\end{table}

\subsection{Computational cost of soft sensors}

The training time refers to the time taken to determine optimal values for the parameters of a soft sensor model. Once the developed soft sensor is implemented online in a distributed control system, it is used to estimate key process or quality variables at regular sampling intervals. The time required to get the estimates is called \emph{soft sensing time}.

Different techniques have various levels of computational complexity, that is, model training time. Since \ac{PCA} \citep{jolliffe2016principal}, \ac{SFA} \citep{wiskott2002slow}, \ac{ICA} \citep{hyvarinen2000independent}, and \ac{FA} \citep{ge2015supervised} can be developed in a single iteration, they require relatively low computational time compared to \acs{LASSO} \citep{tibshirani1996regression}, and \acs{GMM} \citep{yao2018scalable} techniques, which involve using iterative optimization algorithms to determine the model parameters. In general, \ac{ML} methods need more computational time than statistical methods \citep{kadlec2009data}. Further, the computational complexity of \ac{ML} methods is influenced by the factors listed below \citep{kelleher2020fundamentals} :

\begin{itemize}
  \item Amount of training data.
  \item Number of features or input variables.
  \item Type of training algorithm employed.
  \item Number of layers.
  \item Number of neurons (size) in layers.
  \item Type of device used (such as CPU or GPU).
\end{itemize}

The \ac{ELM} is considered the fastest \ac{ML} algorithm because it does not have parameters that need to be learned. The second fastest \ac{ML} method is the \ac{GRNN}, which has a single learnable parameter (spread or width of \iac{RBF}).\footnote{Soft sensors were developed based on industrial data using the \acs{ML} algorithms surveyed in this paper. The soft sensors were tested on an independent validation dataset and the computation time for each soft sensor was recorded. It turned out that, among all of the soft sensors, the \acs{ELM}-based soft sensor took the least computational time to produce estimates. The \acs{GRNN}-based soft sensor was the second fastest one.}
Then come \acp{RT} and \acp{DNNE}, which can be constructed more easily than shallow neural networks like \ac{RBFNN}, \ac{MLP}, \ac{WNN}, and \ac{ANFIS}. As \ac{RBFNN} uses hybrid learning (not hybrid modeling)---unsupervised learning for the middle layer and supervised learning (linear regression) for the last layer---it is usually faster than \ac{MLP}, \ac{ANFIS}, and \ac{WNN}, which use iterative gradient descent algorithms. \Ac{SVM} is the slowest of the kernel-based \ac{ML} methods (\ac{SVM}, \ac{GPR}, and \ac{RVM}). \Acp{BN} rely on the expectation-maximization algorithm to optimize their parameters, which takes a little more training time than \acp{RT}. Dynamic \ac{ML} methods, such as \acp{RNN}, involve more operations than their static \ac{ML} counterparts, so they require more memory and computational power \citep{wang2018deep,goodfellow2016deep}. Similarly, \acp{DNN} often include several layers and hence, contain many parameters. A large amount of training data is necessary to train \acp{DNN}. Therefore, \acp{DNN} are recognized as the most computationally expensive methods of all the data-driven techniques. \par

\subsection{Industry implementation of soft sensors}

In industries, soft sensors are developed by in-house control engineers or third-party contractors (service engineers) from service providers such as Honeywell or Yokogawa. These service providers use their own software to build the soft sensors. When the existing technology used by service providers is inadequate to handle a problem or in-house control engineers have no knowledge of other soft sensing algorithms, the industries provide research funding to universities, research organizations, or startups to develop sophisticated soft sensors to model complex nonlinear processes. The following steps outline how soft sensors are developed and implemented in industries.  \par

\begin{enumerate}
   \item After recognizing a need for a soft sensor application, a team consisting of a panel operator, process engineer, control engineer, and project manager is formed.
	The process engineer prepares a charter to define the core objectives, scope, responsibilities, and timeline of the project.
 This outlines the benefits that the soft sensor project can offer. All the benefits are usually quantified in terms of how much money can be saved. 
 For example, this cost--benefit analysis typically involves weighing the upfront costs---hardware, software, consultants---and continued costs---software licenses, in-house domain experts to handle support and maintenance---against anticipated improved revenue and throughput, as well as reduced cost of the soft sensor.
 Once the team is satisfied with the benefits, the soft sensor project launches.
  \item The next step in executing a soft sensor project involves obtaining process knowledge or expert experience knowledge to identify input variables that have a noteworthy influence on output variables \citep{joe2021dow}. The use of process knowledge or expert experience avoids the inclusion of redundant input variables in soft sensor modeling, leading to reduced model complexity and improved accuracy. In the absence of such knowledge, \ac{ML} algorithms such as LASSO, hybrid LASSO, and ridge regression can be used to identify and remove input variables that have negligible impact on the output variable. 
  \item The third step entails process data collection and preprocessing. The process data are often abundant but poor in information. This is due to significant disturbances, outliers, and missing values. Soft sensors developed using these data may provide incorrect estimates for quality variables. The outliers and missing values from the raw industrial data should be removed to obtain clean data for developing the soft sensor. Although it may not be theoretically rigorous, the usual practice is to detect and delete samples with outliers \citep{chris2013mad}. Missing values are treated in the same fashion. This approach ensures that the clean data are free of outliers and missing values. 
  \item The data collection in industrial settings is often associated with multi-rate sampling. If the sampling frequency of the input variables is higher than that of the output variable, then it is necessary to synchronize the variables. Down-sampling may be used to deal with the multi-rate sampling problem. In the down-sampling approach, samples of the input variables that do not have the respective measurements of the output variable are removed \citep{kadlec2009data}. 
  \item After the process data are preprocessed, they are split into training and validation subsets. The training subset is used to construct a soft sensor model whereas the validation subset is used to evaluate the prediction performance of the soft sensor model. This is called offline validation. The usual practice is to develop a linear model first. If the linear model cannot produce accurate estimates, then more complex statistical or \ac{ML} algorithms are used.
  \item If the soft sensor model delivers satisfactory performance in the offline validation, it is implemented in a distributed control system. Then the performance of the soft sensor is monitored for some time period. If the soft sensor exhibits poor performance, then modifications are made. This is online validation. For offline and online validation, metrics such as the correlation coefficient and root mean squared error are used to quantify the performance of soft sensors \citep{kadlec2009data}.
  In addition, qualitative analysis is considered to see if soft sensor estimates follow the lab data trend. If the soft sensor estimates are poor, the input data are first examined for possible reasons, such as sensor failures, data transmission problems, outliers, plant shutdowns, and plant upsets. Poor estimates can be characterized by low correlation to lab data, estimates out of the operational range, or significant deviation from lab data. If the input data are good, the following strategies are used to get accurate and reliable estimates: 
  \begin{itemize}
  \item Retraining of the soft sensor using the latest data.
  \item Changing the soft sensor modeling algorithm.
  \item Using a different training algorithm.
  \item Changing the parameter initialization method.
  \item Using approaches that can avoid or reduce overfitting.
  \end{itemize}
  Regardless of the type of soft sensor, practicing engineers usually follow the above approach to assess the performance of soft sensors. 
  \item If the online soft sensor consistently provides reasonable results, the soft sensor is used as a measuring device in a control loop. After successfully implementing the soft sensor-based control application, the soft sensor application is handed over to the panel operator. The human-in-the-loop aspect described above is crucial in translating research results into practical applications. 
\end{enumerate}

\subsection{Challenges in soft sensor development}
\label{subsec:soft-sensor-challenges}

Challenges that are often encountered in soft sensor developments are discussed below.

\begin{itemize}
  \item Lack of labeled data is the main challenge that must be dealt with in order to build good soft sensor models. Quality variables are less frequently measured than easily measurable process variables, such as temperature, pressure, flow rate, and level. A sample of a quality variable is collected once every shift (that is, $8$ hours) or $24$ hours. Because of the long sampling interval, an insufficient amount of practical labeled data is available. A soft sensor trained with a limited amount of labeled data may not be able to capture the underlying relationship between the input variables and the output variable. To deal with this problem, a virtual sample generation method may be used to obtain estimated output values for the corresponding input data \citep{li2021virt}. As an alternative, semi-supervised learning may be used to construct the soft sensor. Unsupervised learning algorithms like \ac{PCA}, autoencoders, stacked autoencoders, or deep belief networks can extract features from unlabeled input data. These features are related to the output variable by any data-driven linear or nonlinear model \citep{goodfellow2016deep}.
  \item Operating conditions of the industrial process may change depending on the demand for products, prices of raw materials, and so on. A soft sensor developed using data from one operating condition may not perform well when the operating condition changes. In this situation, multimode soft sensors can be used to get accurate estimates \citep{Shao2018ss}.
  \item Soft sensor maintenance is crucial to continuously attain reasonable estimates, as the performance of an online soft sensor may degrade over time. As a result, estimates obtained by a poorly performing soft sensor do not follow lab data trends. To circumvent this hurdle, the soft sensor is retrained with recent data, and deployed online. A more popular approach to maintain the accuracy of the soft sensor is to adopt a bias updating strategy. In the bias updating strategy, the soft sensor outputs are brought closer to the lab data \citep{Kuilin2015maint}.
\end{itemize}

%===============================================================================

\section{Data-driven and hybrid modeling approaches for optimization and control}
\label{sec:optim&control}
\acbarrier

We revisit data-driven and hybrid modeling in the context of solving optimization and control problems. We further introduce reinforcement learning as an emerging paradigm for solving challenging control tasks. In the same way hybrid modeling represents a spectrum between knowledge-based and data-based modeling, model-based optimization, model predictive control, and reinforcement learning all encompass model-based and model-free methodologies. Naturally, these techniques are also compatible with hybrid modeling approaches, offering new challenges and research opportunities.

\subsection{Model-based optimization}
\label{sec:hybrid-app-1}

A large number of hybrid modeling applications have been geared towards offline process optimization. Here, a hybrid model is appealing because key operational variables in terms of process performance may be included in the mechanistic part of the model. This is to retain sufficient extrapolation while capturing other parts of the process using data-driven techniques, for example, to reduce the computational burden. Local (gradient-based) or stochastic search techniques have traditionally been applied to solve the resulting model-based optimization problems. But a recent trend has been using complete search techniques to overcome convergence to a local optimum and guarantee global optimality in problems with trained machine learning models embedded, such as \ac{MLP} \citep{Schweidtmann2019,Anderson2020,Tsay2021}, \ac{GP} \citep{Schweidtmann2021}, or gradient-boosted trees \citep{Mistry2020}. Applications in chemical engineering include the optimization of simple reactor operations and process flowsheets \citep{Schweidtmann2019} and optimal catalyst selection \citep{Mistry2020}.

It should be noted that developing a data-driven or hybrid model to speed up the optimization of a more fundamental model is akin to conducting a surrogate-based optimization. The latter constitutes an active research area in process flowsheeting, computational fluid dynamics, and molecular dynamics \citep{Biegler2014}. They can be broadly classified into local and global approaches. Global approaches proceed by constructing a surrogate model based on an ensemble of mechanistic simulations before optimizing it, often within an iteration where the surrogate is progressively refined. Several successful implementations rely on \acp{MLP} \citep{Henao2011}, \acp{GP} \citep{Caballero2008,Quirante2015,Kessler2019}, or a combination of various basis functions \citep{Cozad2015,Boukouvala2017} for the surrogate modeling. Practical applications have been for rigorous design of distillation columns \citep{Quirante2015,Kessler2019} and flowsheet or superstructure optimization of chemical processes \citep{Caballero2008,Henao2011}. By contrast, local approaches maintain an accurate surrogate of the mechanistic model within a trust region, whose position and size are adapted iteratively. This procedure entails reconstructing the surrogate model as the trust region moves around. Still, it can offer global convergence guarantees, for example, when the surrogates meet the full linearity property \citep{Conn2009}. Applications of this approach to chemical process optimization include solved-based CO\textsubscript{2} capture \citep{Eason2018} and integrated carbon capture and conversion \citep{Bajaj2018}.

\subsection{Model predictive control and real-time optimization}
\label{sec:hybrid-app-2}

The \ac{RTO} and nonlinear/economic \ac{MPC} methodologies use a process model at their core. 
So far, most successful implementations of \ac{RTO} and \ac{MPC} have relied on mechanistic models \citep{schwenzer2021review,qin2000overview,forbes2015model}. 
But there has been interest in data-driven approaches, which use surrogate models trained on historical data or mechanistic model simulations to drive the optimization. 
The type of surrogate models used in such data-driven \ac{MPC} includes \acp{MLP} \citep{Piche2000,Wu2019} and \acp{GP} \citep{Kocijan2004,Hewing2020}. 
However, comparatively little work has been published on embedding hybrid models into \ac{MPC} to reduce data dependency and infuse physical knowledge for better extrapolation capability \citep{Klimasauskas1998,Zhang2019}. 
\citet{Teixeira2006} applied batch-to-batch optimization to bioprocesses by relying on hybrid models where an adjustable mixture of nonparametric and parametric models represented the cell population subsystem. 
In the \ac{RTO} area, \citet{Cubillos2007} investigated the use of parallel hybrid models with \ac{MLP} embedded on the Williams benchmark plant, but then they had to use stochastic search methods to solve the resulting optimization problems. 
Recently, \citet{Zhang2019} took the extra step of using the same hybrid model simultaneously in the \ac{RTO} and \ac{MPC} layers and demonstrated the benefits for a simulated CSTR and distillation column. Notice that most of these applications consider serial hybrid models with embedded \acp{MLP} to approximate complex nonlinearities in the system. Nevertheless, there is a dearth of industrial or experimental implementations of such technologies to date. \par

\Iac{RTO} methodology that exploits the parallel approach of hybrid semi-parametric modeling at its core is \emph{modifier adaptation} \citep{Chachuat2009}. 
Unlike classical \ac{RTO}, modifier adaptation does not adapt the mechanistic model but adds correction terms---the modifiers---to the cost and constraint functions in the optimization model. 
The original work used process measurements to estimate linear (gradient-based) corrections \citep{Marchetti2009}. 
\citet{Gao2016} proposed combining quadratic regression models trained on available plant data with a nominal mechanistic model to account for curvature information and filter out the process noise.
 Likewise, \citet{Singhal2016} investigated data-driven approaches based on quadratic surrogates as modifiers for the predicted cost and constraint functions and devised an online adaptation strategy for the surrogates inspired by trust-region ideas. Implementations of this \ac{RTO} methodology for industrial systems include load sharing for gas compressors \citep{Milosavljevic2020} and solid-oxide fuel cells \citep{deAvila2019}. \par
 
 More recently, \citet{Ferreira2018} were the first to consider \acp{GP}, trained from past measurement information, as the cost and constraint modifiers. 
 Using nonparametric regression models to describe the plant-model mismatch in \ac{RTO} applications makes sense insofar as the mismatch is generally structural. 
 \Citet{delRio2019,delRio2021} developed this strategy further by introducing modifier-adaptation schemes that rely on trust regions to capture the \acp{GP}' ability to capture the cost and constraint mismatch. 
 Recently, \citet{Petsagkourakis2021} proposed to use co-Kriging to drive the surrogate modeling, where a first (low-fidelity) \ac{GP} emulating the mechanistic process model is integrated within a second (high-fidelity) \ac{GP} that is trained using the process measurements. 
 The benefits of using \acp{GP} in this context lie in their ability to perform real-time uncertainty quantification and allow chance constraints to be satisfied with high confidence. 
 By and large, these developments share many common grounds with surrogate-based optimization techniques (see \cref{sec:hybrid-app-1}), with the added complexity that the process data are noisy and the process optimum might change over time. Finally, it is worth noting that the potential benefits of this \ac{RTO} technology have been mostly investigated through numerical simulation, which cannot substitute for both experimental and industrial validations and should be the subject of future research.

\subsection{Reinforcement learning}
\label{sec:RL}

\Ac{RL} is a class of numerical methods for the data-driven sequential decision-making problem \citep{Sutton2018}. The \ac{RL} agent (algorithm) aims to find an optimal policy, or controller, based on industrial process data collected through interactions with its environment. 

Note that \ac{RL} represents a more general class of techniques from hybrid modeling-based optimization.
Briefly, \ac{RL} includes algorithms for synthesizing control policies without explicit reliance on a model of the process dynamics. 
The supplementary material contains a more precise background on \ac{RL}; readers are also referred to \citet{Sutton2018}.

Finding such a policy requires solving the Bellman equation based on the principle of optimality. However, the equation is often intractable as it ends up with a high-dimensional optimization problem \citep{bertsekas2012dynamic}.
Recent advances in \ac{ML} enable feature analysis of raw sensory-level using \acp{DNN}.
The aid of \acp{DNN} facilitates efficient numerical methods for approximately solving the Bellman equation. Therefore, the scalability of \ac{RL} algorithms has been significantly improved. As a result, so-called deep \ac{RL} is an emerging technology that has shown remarkable performance in real-world and simulated applications such as robotics, autonomous driving, and board games \citep{Levine2016, williams2016aggressive, Silver2017}. 

Deep \ac{RL} has naturally gained attention from the process control community.
In this section, we survey applications of \ac{RL} in process control, and we discuss advances and challenges in \ac{RL} as they potentially pertain to process control applications.

\subsubsection{Reinforcement learning for process control}

With high demands on the performance of process systems, efficient optimization is becoming increasingly essential. 
The ultimate dream goal of any process control system is to develop a controller capable of attaining optimality in large-scale, nonlinear, and hybrid models with constraints, fast online calculation, and adaptation. 
This ideal controller should be amenable to a closed-loop solution
%using a limited number of observations
and robust to online disturbances.

Mathematical programming-based control, such as \ac{MPC} and direct optimization, are popular because they adequately address many of these requirements. \Cref{sec:hybrid-app-1,sec:hybrid-app-2} discuss the mathematical programming paradigm in more detail.
\Ac{RL} has been studied in parallel because it has contrasting features compared to mathematical programming methods \citep{bucsoniu2018reinforcement}. 
According to the review and perspective studies of \citet{shin2019reinforcement, nian2020review, spielberg2019SelfDriving, yoo2021review}, the advantages of \ac{RL} are that: First, a closed-loop state feedback policy can be obtained for generic stochastic control problems, while an open-loop solution is obtained through mathematical programming approaches. 
Most of the computation is done offline by learning the policy through offline data or simulation. 
Assuming that the environment used for offline training is identical to that of the online implementation, the policy is optimal. 
Second, the mathematical programming formulation for stochastic control problems often becomes prohibitively large to be solved within a decision interval. 
On the other hand, uncertainties are implicitly or explicitly quantified by the value or policy functions in \ac{RL} approaches. 
The trained \ac{RL} policy can be implemented with minimal online computation required. 
Third, \ac{RL} is flexible to varying levels of system knowledge, including model-free, partial model-free, and model-based \ac{RL}. 
\Cref{tb:RLMPComparison} summarizes the comparison between \ac{RL} and mathematical programming methods. \par 

\begin{table*}[tbp]
	\caption{A comparison of \acs{RL} and mathematical programming.}
	\begin{center}
		\begin{tabularx}{\linewidth}{LLL}
		\toprule
 & Reinforcement learning & Mathematical programming approaches  \\
 \midrule
Model knowledge & Flexible & Full model \\
Feedback & Trained policy function & Solution of optimization problems \\
Online computation & Negligible & High \\
Offline computation & High & Not required \\ 
Robustness & Backward propagation of uncertain scenario (value-based methods) & Forward propagation of uncertain scenario  \\
Constraint handling & Immature (especially, state variable constraints) & Straightforward  \\
Asymptotic stability & Ultimate upper-boundedness & Asymptotically stable  \\
Scalability & High & Medium \\
Adaptation & Exploitation and exploration can be controlled. However, slow. & Fast. However, performance depends on estimators.\\
		\bottomrule
		\end{tabularx}
	\end{center}
	\label{tb:RLMPComparison}
\end{table*}

Several pioneering pieces of work due to \citet{wilson1997neuro, kaisare2003simulation, peroni2005optimal} proposed applying model-free \ac{RL} to process control problems over discretized state and action spaces. $Q$-learning was implemented for the tracking control of a fed-batch bioreactor \citep{wilson1997neuro} and free-end maximization problem of a fed-batch bioreactor \citep{kaisare2003simulation, peroni2005optimal}. \citet{lee2006choice, lee2009approximate} extended the concept of applying model-free \ac{RL} to dual adaptive control and scheduling problems. It was shown that the approximation of the value function could provide robust control despite the presence of process noise and model changes. \Ac{RL} methods that guarantee robustness in dynamic optimization were later studied in \citet{nosair2010min, yang2013switching}. \par

Some recent applications of \ac{RL} rely on a linear approximator to solve optimal control problems with a continuous state space model \citep{zhu2020scalable, sun2018data, ge2018approximate, kim2018pomdp}. Especially, \citet{zhu2020scalable} applied a model-free \ac{RL} variant called factorial fast-food dynamic policy programming to a Vinyl Acetate monomer process. The algorithm improves scalability by breaking down the exponential size of the action space by action space factorization.
In the meantime, model-free deep \ac{RL} applications have become increasingly studied in the process control field. 
\Cref{tb:RLapps} summarizes some recent work in this area. 
In the remaining sections, we elaborate on the use of deep \ac{RL} in process control.

\begin{table*}[tbp]
	\caption{Model-free deep \acs{RL} applications in process control. Asterisk (*) indicates a model-based modification to the nominal algorithm. \colorbox[gray]{0.9}{Highlighted} rows indicate validation on a physical system.}
	\begin{center}
		\begin{tabularx}{\linewidth}{lll}
		\toprule
 & \acs{RL} algorithm & Application/process  \\
    \midrule
\colorbox[gray]{0.9}{\citet{pandian2018control}} & \acs{DQN}*\citep{Mnih2015} & Quadruple tank system\\
\citet{wang2018NovelApproach} & \acs{PPO}\citep{schulman2017ProximalPolicy} & HVAC control\\
\citet{ma2019continuous} & \acs{DDPG}\citep{lillicrap2015continuous} & Polymerization system\\
\citet{spielberg2019SelfDriving} & \acs{DDPG}\citep{lillicrap2015continuous} & HVAC control\\
\citet{oh2021automatic} & \acs{DQN}\citep{Mnih2015} & Moving bed process\\
\citet{petsagkourakis2020reinforcement} & \acs{REINFORCE}*\citep{williams1992SimpleStatistical} & Fed-batch bioreactor\\
\citet{bao2021DeepReinforcement} & \acs{TD3}*\citep{fujimoto2018addressing} & Setpoint tracking\\
\colorbox[gray]{0.9}{\citet{dogru2021online}} & \acs{A3C}\citep{mnih2016AsynchronousMethods} & Hybrid three-tank system\\
\citet{joshi2021twin} & \acs{TD3}\citep{fujimoto2018addressing} & Transesterification process\\
\citet{mowbray2021UsingProcess} & \acs{REINFORCE}\citep{williams1992SimpleStatistical} & Setpoint tracking\\
\citet{yoo2021reinforcement} & \acs{DDPG}\citep{lillicrap2015continuous} & Polymerization process\\
\colorbox[gray]{0.9}{\citet{lawrence2022deep}} & \acs{TD3}\citep{fujimoto2018addressing} & \acs{PID} tuning\\
\citet{zhu2022benchmark} & \acs{SAC}\citep{haarnoja2018soft} & Polyol process\\
\colorbox[gray]{0.9}{\citet{janjua2023gvfs}} & GVF\citep{sutton2011horde} & Water treatment\\
	\bottomrule
	\end{tabularx}
	\end{center}
	\label{tb:RLapps}
\end{table*}

\subsubsection{Practical implementation of reinforcement learning}
\label{subsec:practicalRL}

\begin{figure}[tbp]
    \centering
    \includegraphics[width=8.4cm]{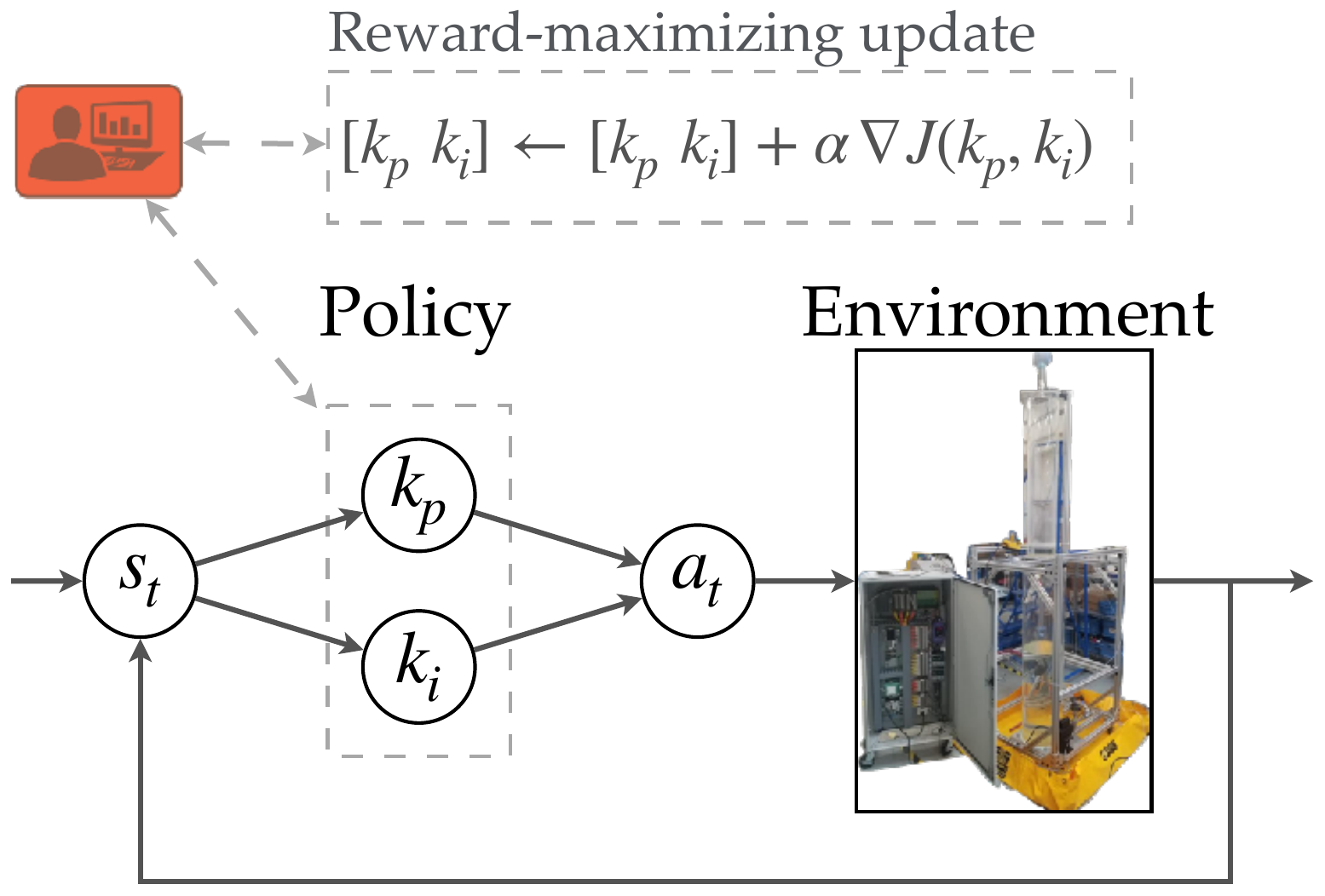}
    \caption{Application of \acs{RL} for tuning \acs{PI} controllers in a lab setting. The policy plays the role of \iacs{PI} controller and receives updates towards improved performance. $J$ is a general long-term cost function and $k_p, k_i$ are controller gains. Adapted from \cite{lawrence2022deep}.}
    \label{fig:RL1}
\end{figure}

One promising application of \ac{RL} is the synthesis of existing control structures \citep{sedighizadeh2008AdaptivePID, carlucho2017IncrementalLearning, shipman2019ReinforcementLearning, kumar2021DiffLoopTuning, lakhani2022StabilitypreservingAutomatic}. 
For example, \ac{PID} controllers constitute the lowest level of control structures, and augmenting these with \ac{RL} methods immediately gives practical results.
\Ac{PID} tuning is a suitable testbed for \ac{RL} applications, as there exists a suite of tuning methods and industrial autotuners to benchmark against \citep{lawrence2022deep}.
\Ac{PID} controllers are also standard in practice, meaning the base layer control is not substituted for a more complex strategy, for example, based on \acp{DNN} (see \cref{fig:RL1}).

Model-free \ac{RL} was applied to schedule a set of \ac{PID} gains obtained \emph{a priori} \citep{lee2006approximate} or from internal model control \citep{brujeni2010dynamic}.
\citet{berger2013neurodynamic} used a model-based \ac{RL} method, called dual heuristic dynamic programming, to compute \ac{PID} gains. 
\citet{nian2020review} applied \ac{DQN} to determine the gains of \ac{PID} controllers and compared the performance with \ac{MPC}. 
\citet{lawrence2022deep} conducted an experimental study on the auto-tuning of \ac{PID} controllers using the \ac{TD3} algorithm. 
\Cref{fig:RL1} depicts a feedback diagram in the \ac{RL} setting: the actor is formulated as \iac{PID} controller for the flow rate to a two-tank system, while the agent processes data on a PC to update the actor-critic parameters.

Another application is to construct hierarchical control structures with \ac{RL} methods.
\citet{shafi2020hierarchical} introduced a two-layer structure for optimizing the bitumen recovery rate of a primary separation vessel. 
A supervisory \ac{RL} agent optimizes the recovery rate, while a low-level \ac{RL} agent computes the interface level actuation. 
\citet{kim2021model} proposed a different type of two-layer structure for a product maximization problem of a fed-batch bioreactor. 
A model-based \ac{RL} agent solves the high-level optimization problem, and \iac{MPC} tracks the trajectory of the high-level optimizer, rejecting real-time disturbances. \par

Several studies make a comparison between \ac{RL} methods based on practical performance criteria.
\citet{wang2019benchmarking} compared 14 model-free and model-based \ac{RL} algorithms based on the following criteria: nominal performance, sample efficiency (total training time, training time per step), robustness against noise, and asymptotic performance. 
\citet{lawrence2022deep} proposed nominal performance, stability, perturbation to the system, initialization, hyperparameters, training duration, practicality, and specialization as key criteria for evaluating \ac{RL} methods for process control problems. 
In addition, \citet{dogru2021online} used the extent of exploration: the ratio of the visited over the total operational state and action spaces. \par

It is worth noting that \ac{RL} implementations on physical systems are sparse.
Some works in process control applications are validated on physical systems \citep{brujeni2010dynamic, syafiie2011ModelfreeControl, pandian2018control, nian2020review, lawrence2022deep, dogru2021online}.
These references tend to focus on \ac{PID} tuning or low-dimensional state/action spaces. A cascaded tank system is also the most common environment.
There are several plausible reasons for the lack of real-world \ac{RL} applications:
The added engineering and software development is not always feasible to accommodate; the algorithmic complexity of \ac{RL} algorithms exacerbates the issue; practical and theoretical problems, such as sample efficiency, convergence, and closed-loop stability, are pressing concerns.
Indeed, most deep \ac{RL} algorithms can achieve impressive final performance on complex tasks, but at the cost of extensive hyperparameter tuning and significant variation between implementations \citep{henderson2018deep}.
In the following section, we highlight a few methods that are geared towards making \ac{RL} more reliable and scalable: Synthesis between model-based and model-free learning; transfer learning and meta-\ac{RL}; offline \ac{RL}.

\subsubsection{Challenges and advances in deep reinforcement learning}

Applying \ac{RL} to industrial settings has many practical, technological, and theoretical challenges.
We refer to \citet{shin2019reinforcement, nian2020review} for further reading. 
Here, we mainly focus on the sample efficiency of \ac{RL} algorithms. 
Sample efficiency refers to the amount of data needed to train an \ac{RL} agent.
The supplementary material contains a more general discussion about \ac{ML} with limited data.
\par 

Classical algorithms for value-based methods, such as $Q$-learning, and policy-based methods, such as \acs{REINFORCE}\footnote{\acl{REINFORCE}}, enjoy theoretical convergence. 
However, convergence can be slow due to high variance in value estimates or limited to the tabular setting or linear function approximation \citep{Sutton2018}. 
Nonetheless, these methods provide the foundation for deep \ac{RL} algorithms. 
Deep \ac{RL} attempts to scale up \ac{RL} methods to high-dimensional problems as a synthesis with the deep learning framework.
The first notable result is an extension of $Q$-learning, named \acp{DQN}, introduced by \citet{Mnih2015}. 
\Acp{DQN} are limited to discrete action spaces but showed impressive results in tasks with high-dimensional sensory input data, such as Atari games.

More recent algorithms, such as the \ac{DDPG} algorithm \citep{lillicrap2015continuous}, allow for continuous action spaces.
Despite the advances made by \ac{DDPG}, it is notoriously difficult to use, for example, due to sensitivity to hyperparameters and overestimation of $Q$-function values \citep{henderson2018deep}. 
This limits the viability of \ac{DDPG} for real-world applications such as process control, as a physical system cannot be extensively probed.
However, the concurrent algorithms, \ac{TD3} \citep{fujimoto2018addressing} and \ac{SAC} \citep{haarnoja2018soft}, built off \ac{DDPG} to improve the overall training robustness and sample efficiency.
%For instance, \citet{haarnoja2018learning} showed that \ac{SAC} could learn a high-performance policy for a physical quadrupedal robot in around two hours.
Despite these advances, model-free \ac{RL} algorithms alone are not sufficiently data-efficient and, therefore, not yet useful in real industrial applications \citep{recht2019tour}. 
In the rest of this section, we identify several areas of \ac{RL} research aimed at this issue.

Although formulating a dynamic model can be a bottleneck in the \ac{RL} algorithm, model-based methods require much fewer interactions with the plant \citep{recht2019tour}. 
Several model-based \ac{RL} algorithms have been developed, focusing on solving the continuous-time counterpart of the Bellman equation called the \ac{HJB} equation. 
Since they aim to solve the \ac{HJB} equation adaptively, the methods are called \ac{ADP} \citep{Prokhorov1997ACD, Lewis2009RL, jiang2014robust}.
\Ac{ADP} algorithms vary with their levels of model utilization, ranging from heuristic dynamic programming, dual heuristic programming, and globalized dual heuristic programming \citep{kim2020gdhp,kim2020convergence}. 
Stochastic optimal control is an extension for handling stochastic differential equations, a continuous-time description for uncertainty. 
\Acf{PI2} is a sampling approach to solving the stochastic \ac{HJB} equation \citep{theodorou2010generalized}. 
\acs{PI2} has shown remarkable data efficiency and performance for robot learning.\par

Another line of work has focused on unifying model-free and model-based approaches \citep{d2020learn, janner2019trust}. 
The main motivation is that model-free algorithms often achieve superior final (asymptotic) performance over model-based approaches but suffer from relatively weak sample complexity.
\Citet{bao2021DeepReinforcement} utilized ideas from \citet{d2020learn} wherein a dynamics model is used to improve the action gradient estimation of the critic network.
 While integrating dynamic models into traditionally model-free algorithms has proved promising, these algorithms are designed to train an agent using online interactions on a system-by-system basis. 
 More general strategies aim to reduce the cost of calibrating \ac{RL} agents to novel environments by utilizing historical datasets, training over many related systems, or transferring previously trained agents to new ones.

\emph{Offline \ac{RL}} (sometimes called \emph{batch \ac{RL}}) aims to learn an optimal policy from historical data alone \citep{levine2020offline}. 
Although off-policy algorithms like \ac{DDPG} can theoretically learn from historical data, online exploration is critical unless constraints are imposed on the learned policy \cite{fujimoto2019off}. 
An offline strategy for pre-training \ac{RL} agents with historical process data, followed by online fine-tuning of the policy, is proposed by \citet{mowbray2021UsingProcess}. 
On the other hand, \emph{\acl{TL}} is a framework for speeding up the training of \ac{RL} agents. 
By pre-training a policy, such as in a simulation environment, one can use this as the initial policy on the true system of interest. 
This idea is demonstrated for batch bioprocess optimization \citep{petsagkourakis2020reinforcement}. 
One can efficiently mitigate plant-model mismatch by fine-tuning the initial policy on the real system. \par

\emph{Meta-learning}, or \emph{learning to learn}, is a \ac{ML} strategy for leveraging prior training experience to learn a new ``task'' quickly \citep{huisman2021SurveyDeep}.
\emph{Meta-\ac{RL}} is a strategy for training a ``meta agent'' to synthesize experience from many related systems to adapt its policy to novel systems rapidly. 
For example, \citet{finn2017model} develop a simple and highly influential algorithm for any neural network architecture that directly optimizes for initial parameters such that they can quickly be adapted to new tasks with a small amount of data, showing superior performance over standard \acl{TL} in classification and \ac{RL} tasks. 
\citet{duan2016rl} propose strategies for learning a latent context variable as part of the meta-policy architecture, thereby capturing the ``task'' structure and enabling the meta-\ac{RL} agent to adapt its policy with new process data. 
This framework is appealing in process control applications because many systems may have a known structure, making training over a distribution of related systems feasible. 
Consequently, this end-to-end framework removes a model identification step during the online implementation of the \ac{RL} agent by leveraging prior training experience. 
Meta-\ac{RL} has also seen recent applications to process control \citep{mcclement2022MetareinforcementLearning}.\par

While significant strides have been made to make these algorithms more sample-efficient, they are not yet practical. 
Motivated by this challenge, we have outlined different ways in which models can be integrated into otherwise model-free algorithms. 
Moreover, meta-\ac{RL}, offline \ac{RL}, and \acl{TL}, while still emerging, are promising avenues for \ac{MPC} applications. 
These areas have tremendous potential for applications that can redefine automation in the process industries.

\section{Discussion}
\label{sec:conclusion}
\acbarrier
% connect soft sensing, modeling, optimization, control, RL, etc...

\begin{table}[tbp]
\caption{Method-application pairs covered in this survey. \cmark: significant emphasis, \xmark: sparse emphasis.}
\begin{center}
\begin{tabular}{lcc}
\toprule
 & Soft sensing & Process control\\
\midrule
Statistical learning & \cmark & \xmark \\
Machine learning & \cmark & \cmark \\
Deep learning & \cmark & \cmark \\
Reinforcement learning
 & \xmark & \cmark\\
 \bottomrule
\end{tabular}
\end{center}
\label{table:summary}
\end{table}

Soft sensing and process control encompass statistical learning, machine learning, deep learning, and reinforcement learning to varying degrees.
\Cref{table:summary} shows the respective high-level prominence in these two application areas.
Although \cref{table:dist_SL} indicates significant interest in the soft sensing literature around deep learning, \cref{table:biao2} shows methods like \acs{PLS} and \acs{SVM} have received the most industrial use.
However, the prominent use of industrial data is still promising.
Meanwhile, our survey of process control indicates a more significant emphasis on deep learning and reinforcement learning in the literature.
Simulation-based studies are commonplace in this context, as discussed in \cref{subsec:practicalRL}.\footnote{Of course, existing implementations of MPC utilize soft sensors and, therefore, statistical/machine learning methods. It is also commonplace in academic studies to augment mathematical programming or reinforcement learning with autoencoders or \acsp{RNN} to learns a state representation \citep{masti2021learning,schlegel2022investigating}.}

\Cref{table:summary} and the above discussion show a duality between sensing and control in the context of machine learning methods.
To fully capture the benefits of modern machine learning methods, a unified framework that encompasses modeling, sensing, and control is required.
Reinforcement learning is well-suited to bridge the gap between sensing and control through a global reward-based objective (rather than treating prediction and control performance as independent goals).
Applications in sensing do not necessarily contradict the model-free nature of reinforcement learning, which is most appealing. 
Rather, this characteristic makes it versatile for processing and optimizing real system data.
To illustrate this point, \citet{xie2023ReinforcementLearning}
propose using reinforcement learning for sensing, even though it has typically been described in the context of control.
Moreover, \citet{nejatbakhshesfahani2023LearningbasedState} utilize reinforcement learning for both state estimation and control under a single closed-loop performance objective.

On the other hand, \cref{sec:RL} discussed the complexity of reinforcement learning algorithms.
More broadly, deep learning and reinforcement learning algorithms are rife with complexity and hyperparameters, making it difficult to parse their fundamental inner workings \citep{henderson2018deep,bras2023ClassicalActorCritic}.
A promising avenue toward unifying sensing and control is distilling reinforcement learning pipelines and reimagining techniques from other branches of machine learning.
Truly robust and powerful methods will follow from such a critical rapprochement of the longstanding statistical learning methods in \cref{fig:biao5} and newer concepts in deep learning and reinforcement learning.
An instance of this aspiration in action is by \citet{eysenbach2023ContrastiveLearning}, where they show a novel use of binary classification and policy iteration is capable of achieving state-of-the-art performance.

\subsection{Conclusions}

Recent advances in \acl{ML} give us renewed optimism for achieving higher levels of automation in the process industries.
To distill this general goal, we have surveyed soft sensing and process control through a practical lens.
Soft sensing represents the most dominant area regarding industrial applications of statistical and \acl{ML} techniques.
On the other hand, considerable research attention has been given to deep learning applications, but with limited industrial successes.
Through synthesizing research trends and industrial requirements, we have strived to enable academics and practitioners alike to develop sophisticated yet practical methods for building better models and controllers.

%===============================================================================

\section*{Acknowledgements}

We are grateful to the anonymous reviewers for their detailed and constructive feedback; their comments significantly improved the quality of this paper.
NPL \& RBG gratefully acknowledge the financial support of the Natural Sciences and Engineering Research Council of Canada (NSERC) and Honeywell Process Solutions.
JML gratefully acknowledges the research facilities for this work provided by the Institute of Engineering Research at Seoul National University.
BC gratefully acknowledges funding by the Engineering and Physical Sciences Research Council (EPSRC) under grants EP/T000414/1 and EP/W003317/1. BH, FA and SKD gratefully acknowledge financial supports from the Natural Sciences and Engineering Research Council of Canada (NSERC) under grants IRCPJ 417793-15 and ALLRP 561080-20.

% \small
% \bibliography{bibliography.bib}             % bib file to produce the bibliography                                                  % with bibtex (preferred)


% \bibliographystyle{elsarticle-num} % getting complaints with this one
%\section*{References}
%\label{sec:bib}
\singlespacing
\renewcommand*{\bibfont}{\footnotesize}

\bibliographystyle{unsrtnat}
\bibliography{bibliography}

\begin{thebibliography}{174}
\providecommand{\natexlab}[1]{#1}
\providecommand{\url}[1]{\texttt{#1}}
\expandafter\ifx\csname urlstyle\endcsname\relax
  \providecommand{\doi}[1]{doi: #1}\else
  \providecommand{\doi}{doi: \begingroup \urlstyle{rm}\Url}\fi

\bibitem[Venkatasubramanian(2019)]{venkatasubramanian2019promise}
Venkat Venkatasubramanian.
\newblock The promise of artificial intelligence in chemical engineering: Is it
  here, finally?
\newblock \emph{AIChE Journal}, 65\penalty0 (2):\penalty0 466--478, 2019.

\bibitem[Chiang et~al.(2000)Chiang, Russell, and Braatz]{chiang2000fault}
Leo~H Chiang, Evan~L Russell, and Richard~D Braatz.
\newblock \emph{Fault detection and diagnosis in industrial systems}.
\newblock Springer Science \& Business Media, 2000.

\bibitem[Qin and Chiang(2019)]{qin2019advances}
S~Joe Qin and Leo~H Chiang.
\newblock Advances and opportunities in machine learning for process data
  analytics.
\newblock \emph{Computers \& Chemical Engineering}, 126:\penalty0 465--473,
  2019.

\bibitem[Ge et~al.(2017)Ge, Song, Ding, and Huang]{ge2017DataMining}
Zhiqiang Ge, Zhihuan Song, Steven~X Ding, and Biao Huang.
\newblock Data mining and analytics in the process industry: The role of
  machine learning.
\newblock \emph{IEEE Access}, 5:\penalty0 20590--20616, 2017.

\bibitem[Shin et~al.(2019)Shin, Badgwell, Liu, and Lee]{shin2019reinforcement}
Joohyun Shin, Thomas~A Badgwell, Kuang-Hung Liu, and Jay~H Lee.
\newblock {Reinforcement Learning--Overview of recent progress and implications
  for process control}.
\newblock \emph{Computers \& Chemical Engineering}, 127:\penalty0 282--294,
  2019.

\bibitem[Spielberg et~al.(2019)Spielberg, Tulsyan, Lawrence, Loewen, and
  Gopaluni]{spielberg2019SelfDriving}
Steven Spielberg, Aditya Tulsyan, Nathan~P. Lawrence, Philip~D. Loewen, and
  R.~Bhushan Gopaluni.
\newblock Toward self-driving processes: {{A}} deep reinforcement learning
  approach to control.
\newblock \emph{AIChE Journal}, 65, 2019.

\bibitem[Tulsyan et~al.(2020)Tulsyan, Wang, Schorner, Khodabandehlou, Coufal,
  and Undey]{tulsyan2019automatic}
Aditya Tulsyan, Tony Wang, Gregg Schorner, Hamid Khodabandehlou, Myra Coufal,
  and Cenk Undey.
\newblock Automatic real-time calibration, assessment, and maintenance of
  generic {Raman} models for online monitoring of cell culture processes.
\newblock \emph{Biotechnology and Bioengineering}, 117\penalty0 (2):\penalty0
  404--416, 2020.

\bibitem[Nian et~al.(2020)Nian, Liu, and Huang]{nian2020review}
Rui Nian, Jinfeng Liu, and Biao Huang.
\newblock A review on reinforcement learning: Introduction and applications in
  industrial process control.
\newblock \emph{Computers \& Chemical Engineering}, page 106886, 2020.

\bibitem[Bi et~al.(2022)Bi, Qin, Wu, Zheng, and Zhao]{bi2022one}
Xiaotian Bi, Ruoshi Qin, Deyang Wu, Shaodong Zheng, and Jinsong Zhao.
\newblock One step forward for smart chemical process fault detection and
  diagnosis.
\newblock \emph{Computers \& Chemical Engineering}, 164:\penalty0 107884, 2022.

\bibitem[Gamer et~al.(2020)Gamer, Hoernicke, Kloepper, Bauer, and
  Isaksson]{gamer2020autonomous}
Thomas Gamer, Mario Hoernicke, Benjamin Kloepper, Reinhard Bauer, and Alf~J
  Isaksson.
\newblock The autonomous industrial plant--future of process engineering,
  operations and maintenance.
\newblock \emph{Journal of Process Control}, 88:\penalty0 101--110, 2020.

\bibitem[Gopaluni et~al.(2020)Gopaluni, Tulsyan, Chachuat, Huang, Lee, Amjad,
  Damarla, Kim, and Lawrence]{gopaluni2020modern}
R.~Bhushan Gopaluni, Aditya Tulsyan, Benoit Chachuat, Biao Huang, Jong~Min Lee,
  Faraz Amjad, Seshu~Kumar Damarla, Jong~Woo Kim, and Nathan~P. Lawrence.
\newblock Modern machine learning tools for monitoring and control of
  industrial processes: A survey.
\newblock \emph{IFAC-PapersOnLine}, 53\penalty0 (2):\penalty0 218--229, 2020.

\bibitem[Sansana et~al.(2021)Sansana, Joswiak, Castillo, Wang, Rendall, Chiang,
  and Reis]{sansana2021RecentTrends}
Joel Sansana, Mark~N Joswiak, Ivan Castillo, Zhenyu Wang, Ricardo Rendall,
  Leo~H Chiang, and Marco~S Reis.
\newblock Recent trends on hybrid modeling for industry 4.0.
\newblock \emph{Computers \& Chemical Engineering}, 151:\penalty0 107365, 2021.

\bibitem[Hastie et~al.(2009)Hastie, Tibshirani, and
  Friedman]{hastie2009elements}
Trevor Hastie, Robert Tibshirani, and Jerome Friedman.
\newblock \emph{The Elements of Statistical Learning: Data Mining, Inference
  and Prediction}.
\newblock Springer Series in Statistics, New York, 2nd edition, 2009.

\bibitem[Goodfellow et~al.(2016)Goodfellow, Bengio, and
  Courville]{goodfellow2016deep}
Ian Goodfellow, Yoshua Bengio, and Aaron Courville.
\newblock \emph{Deep learning}.
\newblock MIT press, 2016.

\bibitem[von Stosch et~al.(2014)von Stosch, Oliveira, Peres, and Feyo~de
  Azevedo]{VonStosch2014}
Moritz von Stosch, Rui Oliveira, Joana Peres, and Sebastião Feyo~de Azevedo.
\newblock Hybrid semi-parametric modeling in process systems engineering:
  {P}ast, present and future.
\newblock \emph{Computers \& Chemical Engineering}, 60:\penalty0 86 -- 101,
  2014.

\bibitem[Peherstorfer et~al.(2018)Peherstorfer, Willcox, and
  Gunzburger]{Peherstorfer2018}
Benjamin Peherstorfer, Karen Willcox, and Max Gunzburger.
\newblock Survey of multifidelity methods in uncertainty propagation,
  inference, and optimization.
\newblock \emph{SIAM Review}, 60\penalty0 (3):\penalty0 550--591, 2018.

\bibitem[Psichogios and Ungar(1992)]{Psichogios1992}
Dimitris~C. Psichogios and Lyle~H. Ungar.
\newblock A hybrid neural network-first principles approach to process
  modeling.
\newblock \emph{AIChE Journal}, 38\penalty0 (10):\penalty0 1499--1511, 1992.

\bibitem[Su et~al.(1992)Su, Bhat, Minderman, and McAvoy]{Su1992}
Hong-Te Su, N.~Bhat, P.~A. Minderman, and T.~J. McAvoy.
\newblock Integrating neural networks with first principles models for dynamic
  modeling.
\newblock \emph{IFAC Proceedings Volumes}, 25\penalty0 (5):\penalty0 327--332,
  1992.

\bibitem[Thompson and Kramer(1994)]{Thompson1994}
Michael~L. Thompson and Mark~A. Kramer.
\newblock Modeling chemical processes using prior knowledge and neural
  networks.
\newblock \emph{AIChE Journal}, 40\penalty0 (8):\penalty0 1328--1340, 1994.

\bibitem[Chen et~al.(2000)Chen, Bernard, Bastin, and Angelov]{Chen2000}
L.~Chen, O.~Bernard, G.~Bastin, and P.~Angelov.
\newblock Hybrid modelling of biotechnological processes using neural networks.
\newblock \emph{Control Engineering Practice}, 8\penalty0 (7):\penalty0
  821--827, 2000.

\bibitem[Solle et~al.(2017)Solle, Hitzmann, Herwig, Pereira~Remelhe, Ulonska,
  Wuerth, Prata, and Steckenreiter]{Solle2017}
Dörte Solle, Bernd Hitzmann, Christoph Herwig, Manuel Pereira~Remelhe, Sophia
  Ulonska, Lynn Wuerth, Adrian Prata, and Thomas Steckenreiter.
\newblock Between the poles of data-driven and mechanistic modeling for process
  operation.
\newblock \emph{Chemie Ingenieur Technik}, 89\penalty0 (5):\penalty0 542--561,
  2017.

\bibitem[Schuppert and Mrziglod(2018)]{Schuppert2018}
A.~Schuppert and T.~Mrziglod.
\newblock Hybrid model identification and discrimination with practical
  examples from the chemical industry.
\newblock In J.~Glassey and M.~von Stosch, editors, \emph{Hybrid Modeling in
  Process Industries}, pages 63--88, Boca Raton, 2018. CRC Press.

\bibitem[Zendehboudi et~al.(2018)Zendehboudi, Rezaei, and
  Lohi]{Zendehboudi2018}
Sohrab Zendehboudi, Nima Rezaei, and Ali Lohi.
\newblock Applications of hybrid models in chemical, petroleum, and energy
  systems: A systematic review.
\newblock \emph{Applied Energy}, 228:\penalty0 2539--2566, 2018.

\bibitem[Ahmad et~al.(2020)Ahmad, Ayub, Kano, and Cheema]{Ahmad2020}
I.~Ahmad, A.~Ayub, M.~Kano, and I.I. Cheema.
\newblock Gray-box soft sensors in process industry: Current practice, and
  future prospects in era of big data.
\newblock \emph{Processes}, 8\penalty0 (2):\penalty0 243, 2020.

\bibitem[Bradley et~al.(2022)Bradley, Kim, Kilwein, Blakely, Eydenberg, Jalvin,
  Laird, and Boukouvala]{bradley2022}
William Bradley, Jinhyeun Kim, Zachary Kilwein, Logan Blakely, Michael
  Eydenberg, Jordan Jalvin, Carl Laird, and Fani Boukouvala.
\newblock Perspectives on the integration between first-principles and
  data-driven modeling.
\newblock \emph{Computers \& Chemical Engineering}, 166:\penalty0 107898, 2022.

\bibitem[Agarwal(1997)]{Agarwal1997}
Mukul Agarwal.
\newblock Combining neural and conventional paradigms for modelling, prediction
  and control.
\newblock \emph{International Journal of Systems Science}, 28\penalty0
  (1):\penalty0 65--81, 1997.

\bibitem[Bangi and Kwon(2020)]{Bangi2020}
Mohammed Saad~Faizan Bangi and Joseph Sang-Il Kwon.
\newblock Deep hybrid modeling of chemical process: Application to hydraulic
  fracturing.
\newblock \emph{Computers \& Chemical Engineering}, 134:\penalty0 106696, 2020.

\bibitem[Chaffart and Ricardez-Sandoval(2018)]{Chaffart2018}
Donovan Chaffart and Luis~A. Ricardez-Sandoval.
\newblock Optimization and control of a thin film growth process: A hybrid
  first principles/artificial neural network based multiscale modelling
  approach.
\newblock \emph{Computers \& Chemical Engineering}, 119:\penalty0 465--479,
  2018.

\bibitem[Ghosh et~al.(2019)Ghosh, Hermonat, Mhaskar, Snowling, and
  Goel]{Ghosh2019}
D.~Ghosh, E.~Hermonat, P.~Mhaskar, S.~Snowling, and R.~Goel.
\newblock Hybrid modeling approach integrating first-principles models with
  subspace identification.
\newblock \emph{Industrial \& Engineering Chemistry Research}, 58\penalty0
  (30):\penalty0 13533--13543, 2019.

\bibitem[Lopez et~al.(2020)Lopez, Udugama, Thomsen, Roslander, Junicke,
  Mauricio-Iglesias, and Gernaey]{Lopez2020}
Pau~Cabaneros Lopez, Isuru~A. Udugama, Sune~T. Thomsen, Christian Roslander,
  Helena Junicke, Miguel Mauricio-Iglesias, and Krist~V. Gernaey.
\newblock Towards a digital twin: a hybrid data-driven and mechanistic digital
  shadow to forecast the evolution of lignocellulosic fermentation.
\newblock \emph{Biofuels, Bioproducts \& Biorefining}, 14\penalty0
  (5):\penalty0 1046--1060, 2020.

\bibitem[Zhang et~al.(2019{\natexlab{a}})Zhang, Del Rio-Chanona,
  Petsagkourakis, and Wagner]{Zhang2019b}
Dongda Zhang, Ehecatl~Antonio Del Rio-Chanona, Panagiotis Petsagkourakis, and
  Jonathan Wagner.
\newblock Hybrid physics-based and data-driven modeling for bioprocess online
  simulation and optimization.
\newblock \emph{Biotechnology \& Bioengineering}, 116\penalty0 (11):\penalty0
  2919--2930, 2019{\natexlab{a}}.

\bibitem[Duarte et~al.(2004)Duarte, Saraiva, and Pantelides]{Duarte2004}
Belmiro Duarte, P.~M. Saraiva, and C.~C. Pantelides.
\newblock Combined mechanistic and empirical modelling.
\newblock \emph{International Journal of Chemical Reactor Engineering},
  2\penalty0 (1), 2004.

\bibitem[de~Prada et~al.(2018)de~Prada, Hose, Gutierrez, and
  Pitarch]{dePrada2018}
C.~de~Prada, D.~Hose, G.~Gutierrez, and J.~L. Pitarch.
\newblock Developing grey-box dynamic process models.
\newblock \emph{IFAC-PapersOnLine}, 51\penalty0 (2):\penalty0 523--528, 2018.

\bibitem[Hesterberg et~al.(2008)Hesterberg, Choi, Meier, and
  Fraley]{Hesterberg2008}
Tim Hesterberg, Nam~Hee Choi, Lukas Meier, and Chris Fraley.
\newblock Least angle and $\ell_1$ penalized regression: A review.
\newblock \emph{Statistics Surveys}, 2:\penalty0 61--93, 2008.

\bibitem[Willis and von Stosch(2017)]{Willis2017}
Mark~J. Willis and Moritz von Stosch.
\newblock Simultaneous parameter identification and discrimination of the
  nonparametric structure of hybrid semi-parametric models.
\newblock \emph{Computers \& Chemical Engineering}, 104:\penalty0 366--376,
  2017.

\bibitem[Zhang et~al.(2020)Zhang, Savage, and Cho]{Zhang2020}
Dongda Zhang, Thomas~R. Savage, and Bovinille~A. Cho.
\newblock Combining model structure identification and hybrid modelling for
  photo-production process predictive simulation and optimisation.
\newblock \emph{Biotechnology \& Bioengineering}, 117\penalty0 (11):\penalty0
  3356--3367, 2020.

\bibitem[Brunton et~al.(2016)Brunton, Proctor, and Kutz]{Brunton2016}
Steven~L. Brunton, Joshua~L. Proctor, and J.~Nathan Kutz.
\newblock Discovering governing equations from data by sparse identification of
  nonlinear dynamical systems.
\newblock \emph{Proceedings of the National Academy of Sciences}, 113\penalty0
  (15):\penalty0 3932--3937, 2016.

\bibitem[Wilson and Sahinidis(2017)]{Wilson2017}
Zachary~T. Wilson and Nikolaos~V. Sahinidis.
\newblock The {ALAMO} approach to machine learning.
\newblock \emph{Computers \& Chemical Engineering}, 106:\penalty0 785--795,
  2017.

\bibitem[Cozad et~al.(2015)Cozad, Sahinidis, and Miller]{Cozad2015}
Alison Cozad, Nikolaos~V. Sahinidis, and David~C. Miller.
\newblock A combined first-principles and data-driven approach to model
  building.
\newblock \emph{Computers \& Chemical Engineering}, 73:\penalty0 116--127,
  2015.

\bibitem[Nauta et~al.(2007)Nauta, Weiland, Backx, and Jokic]{Nauta2007}
K.~M. Nauta, S.~Weiland, A.~C. Backx, and A.~Jokic.
\newblock Approximation of fast dynamics in kinetic networks using non-negative
  polynomials.
\newblock In \emph{2007 IEEE International Conference on Control Applications},
  pages 1144--1149, 2007.

\bibitem[Pitarch et~al.(2016)Pitarch, Sala, and de~Prada]{Pitarch2019}
J.~L. Pitarch, A.~Sala, and C.~de~Prada.
\newblock A systematic grey-box modeling methodology via data reconciliation
  and {SOS} constrained regression.
\newblock \emph{Processes}, 7\penalty0 (3):\penalty0 170, 2016.

\bibitem[Raissi et~al.(2019)Raissi, Perdikaris, and Karniadakis]{Raissi2019}
M.~Raissi, P.~Perdikaris, and G.~E. Karniadakis.
\newblock Physics-informed neural networks: A deep learning framework for
  solving forward and inverse problems involving nonlinear partial differential
  equations.
\newblock \emph{Journal of Computational Physics}, 378:\penalty0 686 -- 707,
  2019.

\bibitem[Carey and Finlayson(1975)]{Carey1975}
G.~F. Carey and B.~A. Finlayson.
\newblock Orthogonal collocation on finite elements.
\newblock \emph{Chemical Engineering Science}, 30\penalty0 (5):\penalty0
  587--596, 1975.

\bibitem[Biegler(2007)]{biegler2007overview}
Lorenz~T Biegler.
\newblock An overview of simultaneous strategies for dynamic optimization.
\newblock \emph{{Chemical Engineering and Processing: Process
  Intensification}}, 46\penalty0 (11):\penalty0 1043--1053, 2007.

\bibitem[Liu et~al.(2018)Liu, Cai, and Ong]{Liu2018}
H.~Liu, J.~Cai, and Y.~S. Ong.
\newblock Remarks on multi-output {Gaussian} process regression.
\newblock \emph{Knowledge-Based Systems}, 144:\penalty0 102--121, 2018.

\bibitem[Le~Gratiet and Garnier(2014)]{LeGratiet2014}
L.~Le~Gratiet and J.~Garnier.
\newblock Recursive co-kriging model for design of computer experiments with
  multiple levels of fidelity.
\newblock \emph{International Journal for Uncertainty Quantification},
  4\penalty0 (5):\penalty0 365--386, 2014.

\bibitem[Perdikaris et~al.(2017)Perdikaris, Raissi, Damianou, Lawrence, and
  Karniadakis]{Perdikaris2017}
P.~Perdikaris, M.~Raissi, A.~Damianou, N.~D. Lawrence, and G.~E. Karniadakis.
\newblock Nonlinear information fusion algorithms for data-efficient
  multi-fidelity modelling.
\newblock \emph{Proceedings of the Royal Society A: Mathematical, Physical and
  Engineering Sciences}, 473\penalty0 (2198), 2017.

\bibitem[Cutajar et~al.(2018)Cutajar, Pullin, Damianou, Lawrence, and
  Gonz\'alez]{Cutajar2018}
K.~Cutajar, M.~Pullin, A.~Damianou, N.~Lawrence, and J.~Gonz\'alez.
\newblock Deep {G}aussian processes for multi-fidelity modeling.
\newblock \emph{Advances in Neural Information Processing Systems}, 32, 2018.

\bibitem[Ahmadi and Khadir(2020)]{Ahmadi2020}
Amir~Ali Ahmadi and Bachir~El Khadir.
\newblock Learning dynamical systems with side information, 2020.

\bibitem[Khatibisepehr et~al.(2013)Khatibisepehr, Huang, and
  Khare]{khatibisepehr2013design}
Shima Khatibisepehr, Biao Huang, and Swanand Khare.
\newblock Design of inferential sensors in the process industry: A review of
  {Bayesian} methods.
\newblock \emph{Journal of Process Control}, 23\penalty0 (10):\penalty0
  1575--1596, 2013.

\bibitem[Fortuna et~al.(2007)Fortuna, Graziani, Rizzo, and
  Xibilia]{fortuna2007soft}
Luigi Fortuna, Salvatore Graziani, Alessandro Rizzo, and Maria~Gabriella
  Xibilia.
\newblock \emph{Soft sensors for monitoring and control of industrial
  processes}.
\newblock Springer Science \& Business Media, 2007.

\bibitem[Chu et~al.(2019)Chu, Zhao, Yao, Chen, and Wang]{chu2019transfer}
Fei Chu, Xu~Zhao, Yuan Yao, Tao Chen, and Fuli Wang.
\newblock Transfer learning for batch process optimal control using {LV-PTM}
  and adaptive control strategy.
\newblock \emph{Journal of Process Control}, 81:\penalty0 197--208, 2019.

\bibitem[Kadlec et~al.(2009)Kadlec, Gabrys, and Strandt]{kadlec2009data}
Petr Kadlec, Bogdan Gabrys, and Sibylle Strandt.
\newblock Data-driven soft sensors in the process industry.
\newblock \emph{Computers \& Chemical Engineering}, 33\penalty0 (4):\penalty0
  795--814, 2009.

\bibitem[Jolliffe and Cadima(2016)]{jolliffe2016principal}
Ian~T Jolliffe and Jorge Cadima.
\newblock Principal component analysis: a review and recent developments.
\newblock \emph{Philosophical Transactions of the Royal Society A:
  Mathematical, Physical and Engineering Sciences}, 374\penalty0
  (2065):\penalty0 20150202, 2016.

\bibitem[Wiskott and Sejnowski(2002)]{wiskott2002slow}
Laurenz Wiskott and Terrence~J Sejnowski.
\newblock Slow feature analysis: Unsupervised learning of invariances.
\newblock \emph{Neural computation}, 14\penalty0 (4):\penalty0 715--770, 2002.

\bibitem[Hyv{\"a}rinen and Oja(2000)]{hyvarinen2000independent}
Aapo Hyv{\"a}rinen and Erkki Oja.
\newblock Independent component analysis: algorithms and applications.
\newblock \emph{Neural networks}, 13\penalty0 (4-5):\penalty0 411--430, 2000.

\bibitem[Ge(2015)]{ge2015supervised}
Zhiqiang Ge.
\newblock Supervised latent factor analysis for process data regression
  modeling and soft sensor application.
\newblock \emph{IEEE Transactions on Control Systems Technology}, 24\penalty0
  (3):\penalty0 1004--1011, 2015.

\bibitem[Tibshirani(1996)]{tibshirani1996regression}
Robert Tibshirani.
\newblock Regression shrinkage and selection via the lasso.
\newblock \emph{Journal of the Royal Statistical Society: Series B
  (Methodological)}, 58\penalty0 (1):\penalty0 267--288, 1996.

\bibitem[Yao and Ge(2018)]{yao2018scalable}
Le~Yao and Zhiqiang Ge.
\newblock Scalable semisupervised {GMM} for big data quality prediction in
  multimode processes.
\newblock \emph{IEEE Transactions on Industrial Electronics}, 66\penalty0
  (5):\penalty0 3681--3692, 2018.

\bibitem[Kelleher et~al.(2020)Kelleher, Mac~Namee, and
  D'arcy]{kelleher2020fundamentals}
John~D Kelleher, Brian Mac~Namee, and Aoife D'arcy.
\newblock \emph{Fundamentals of machine learning for predictive data analytics:
  algorithms, worked examples, and case studies}.
\newblock MIT press, 2020.

\bibitem[Wang et~al.(2018{\natexlab{a}})Wang, Gopaluni, Chen, and
  Song]{wang2018deep}
Kai Wang, Bhushan Gopaluni, Junghui Chen, and Zhihuan Song.
\newblock Deep learning of complex batch process data and its application on
  quality prediction.
\newblock \emph{IEEE Transactions on Industrial Informatics},
  2018{\natexlab{a}}.

\bibitem[Qin et~al.(2021)Qin, Guo, Li, Chiang, Castillo, Braun, and
  Wang]{joe2021dow}
S~Joe Qin, Siyi Guo, Zheyu Li, Leo~H Chiang, Ivan Castillo, Birgit Braun, and
  Zhenyu Wang.
\newblock Integration of process knowledge and statistical learning for the
  {Dow} data challenge problem.
\newblock \emph{Computers \& Chemical Engineering}, 153:\penalty0 107451, 2021.

\bibitem[Leys et~al.(2013)Leys, Ley, Klein, Bernard, and Licata]{chris2013mad}
Christophe Leys, Christophe Ley, Olivier Klein, Philippe Bernard, and Laurent
  Licata.
\newblock Detecting outliers: Do not use standard deviation around the mean,
  use absolute deviation around the median.
\newblock \emph{Journal of experimental social psychology}, 49\penalty0
  (4):\penalty0 764--766, 2013.

\bibitem[Ling~Li and Huang(2021)]{li2021virt}
Yalin~Wang Ling~Li, Seshu Kumar~Damarla and Biao Huang.
\newblock A {Gaussian} mixture model based virtual sample generation approach
  for small datasets in industrial processes.
\newblock \emph{Information Sciences}, 581:\penalty0 262--277, 2021.

\bibitem[Weiming~Shao and Yao(2018)]{Shao2018ss}
Zhihuan~Song Weiming~Shao and Le~Yao.
\newblock Soft sensor development for multimode processes based on
  semisupervised {Gaussian} mixture models.
\newblock \emph{IFAC PapersOnline}, 51:\penalty0 614--619, 2018.

\bibitem[Chen et~al.(2015)Chen, Castillo, Chiang, and Yu]{Kuilin2015maint}
Kuilin Chen, Ivan Castillo, Leo~H Chiang, and Jie Yu.
\newblock Soft sensor model maintenance: A case study in industrial processes.
\newblock \emph{IFAC-PapersOnLine}, 48\penalty0 (8):\penalty0 427--432, 2015.

\bibitem[Schweidtmann and Mitsos(2019)]{Schweidtmann2019}
Artur~M. Schweidtmann and Alexander Mitsos.
\newblock Deterministic global optimization with artificial neural networks
  embedded.
\newblock \emph{Journal of Optimization Theory \& Applications}, 180\penalty0
  (3):\penalty0 925--948, 2019.

\bibitem[Anderson et~al.(2020)Anderson, Huchette, Ma, Tjandraatmadja, and
  Vielma]{Anderson2020}
R.~Anderson, J.~Huchette, W.~Ma, C.~Tjandraatmadja, and J.~P. Vielma.
\newblock Strong mixed-integer programming formulations for trained neural
  networks.
\newblock \emph{Mathematical Programming}, 183:\penalty0 3--39, 2020.

\bibitem[Tsay et~al.(2021)Tsay, Kronqvist, Thebelt, and Misener]{Tsay2021}
C.~Tsay, J.~Kronqvist, A.~Thebelt, and R.~Misener.
\newblock Partition-based formulations for mixed-integer optimization of
  trained {ReLU} neural networks.
\newblock \emph{Advances in Neural Information Processing Systems},
  34:\penalty0 3068--3080, 2021.

\bibitem[Schweidtmann et~al.(2021)Schweidtmann, Bongartz, Grothe, Kerkenhoff,
  Lin, Najman, and Mitsos]{Schweidtmann2021}
Artur~M. Schweidtmann, Dominik Bongartz, Daniel Grothe, Tim Kerkenhoff,
  Xiaopeng Lin, Jaromil Najman, and Alexander Mitsos.
\newblock Deterministic global optimization with {Gaussian} processes embedded.
\newblock \emph{Mathematical Programming Computation}, 13:\penalty0 553–581,
  2021.

\bibitem[Mistry et~al.(2021)Mistry, Letsios, Krennrich, Lee, and
  Misener]{Mistry2020}
M.~Mistry, D.~Letsios, G.~Krennrich, R.~M. Lee, and R.~Misener.
\newblock Mixed-integer convex nonlinear optimization with gradient-boosted
  trees embedded.
\newblock \emph{INFORMS Journal on Computing}, 33\penalty0 (3):\penalty0
  1103--1119, 2021.

\bibitem[Biegler et~al.(2014)Biegler, Lang, and Lin]{Biegler2014}
Lorenz~T. Biegler, Yi-dong Lang, and Weijie Lin.
\newblock Multi-scale optimization for process systems engineering.
\newblock \emph{Computers \& Chemical Engineering}, 60:\penalty0 17--30, 2014.

\bibitem[Henao and Maravelias(2011)]{Henao2011}
Carlos~A. Henao and Christos~T. Maravelias.
\newblock Surrogate-based superstructure optimization framework.
\newblock \emph{AIChE Journal}, 57\penalty0 (5):\penalty0 1216--1232, 2011.

\bibitem[Caballero and Grossmann(2008)]{Caballero2008}
Jos\'{e}~A. Caballero and Ignacio~E. Grossmann.
\newblock An algorithm for the use of surrogate models in modular flowsheet
  optimization.
\newblock \emph{AIChE Journal}, 54\penalty0 (10):\penalty0 2633--2650, 2008.

\bibitem[Quirante et~al.(2015)Quirante, Javaloyes, and Caballero]{Quirante2015}
Natalia Quirante, Juan Javaloyes, and Jos\'{e}. Caballero.
\newblock Rigorous design of distillation columns using surrogate models based
  on kriging interpolation.
\newblock \emph{AIChE Journal}, 61\penalty0 (7):\penalty0 2169--2187, 2015.

\bibitem[Ke{\ss}ler et~al.(2019)Ke{\ss}ler, Kunde, McBride, Mertens, Michaels,
  Sundmacher, and Kienle]{Kessler2019}
Tobias Ke{\ss}ler, Christian Kunde, Kevin McBride, Nick Mertens, Dennis
  Michaels, Kai Sundmacher, and Achim Kienle.
\newblock Global optimization of distillation columns using explicit and
  implicit surrogate models.
\newblock \emph{Chemical Engineering Science}, 197:\penalty0 235--245, 2019.

\bibitem[Boukouvala and Floudas(2017)]{Boukouvala2017}
Fani Boukouvala and Christodoulos~A. Floudas.
\newblock {ARGONAUT: AlgoRithms for Global Optimization of coNstrAined grey-box
  compUTational problems}.
\newblock \emph{Optimization Letters}, 11\penalty0 (5):\penalty0 895--913,
  2017.

\bibitem[Conn et~al.(2009)Conn, Scheinberg, and Vicente]{Conn2009}
Andrew~R. Conn, Katya Scheinberg, and Luis~N. Vicente.
\newblock \emph{Introduction to Derivative-Free Optimization}.
\newblock MOS-SIAM Series on Optimization, 2009.

\bibitem[Eason and Biegler(2018)]{Eason2018}
John~P. Eason and Lorenz~T. Biegler.
\newblock Advanced trust region optimization strategies for glass box/black box
  models.
\newblock \emph{AIChE Journal}, 64\penalty0 (11):\penalty0 3934--3943, 2018.

\bibitem[Bajaj et~al.(2018)Bajaj, Iyer, and Hasan]{Bajaj2018}
Ishan Bajaj, Shachit~S. Iyer, and M.~M.~Faruque Hasan.
\newblock A trust region-based two phase algorithm for constrained black-box
  and grey-box optimization with infeasible initial point.
\newblock \emph{Computers \& Chemical Engineering}, 116:\penalty0 306--321,
  2018.

\bibitem[Schwenzer et~al.(2021)Schwenzer, Ay, Bergs, and
  Abel]{schwenzer2021review}
Max Schwenzer, Muzaffer Ay, Thomas Bergs, and Dirk Abel.
\newblock Review on model predictive control: An engineering perspective.
\newblock \emph{The International Journal of Advanced Manufacturing
  Technology}, 117\penalty0 (5-6):\penalty0 1327--1349, 2021.

\bibitem[Qin and Badgwell(2000)]{qin2000overview}
S~Joe Qin and Thomas~A Badgwell.
\newblock An overview of nonlinear model predictive control applications.
\newblock \emph{Nonlinear model predictive control}, pages 369--392, 2000.

\bibitem[Forbes et~al.(2015)Forbes, Patwardhan, Hamadah, and
  Gopaluni]{forbes2015model}
Michael~G Forbes, Rohit~S Patwardhan, Hamza Hamadah, and R~Bhushan Gopaluni.
\newblock Model predictive control in industry: Challenges and opportunities.
\newblock \emph{IFAC-PapersOnLine}, 48\penalty0 (8):\penalty0 531--538, 2015.

\bibitem[Piche et~al.(2000)Piche, Sayyar-Rodsari, Johnson, and
  Gerules]{Piche2000}
Stephen Piche, Bijan Sayyar-Rodsari, Doug Johnson, and Mark Gerules.
\newblock {Nonlinear model predictive control using neural networks}.
\newblock \emph{IEEE Control Systems Magazine}, 20\penalty0 (3):\penalty0
  53--62, 2000.

\bibitem[Wu et~al.(2019)Wu, Tran, Rincon, and Christofides]{Wu2019}
Zhe Wu, Anh Tran, David Rincon, and Panagiotis~D. Christofides.
\newblock Machine learning-based predictive control of nonlinear processes.
  {P}art {I}: {T}heory.
\newblock \emph{AIChE Journal}, 65\penalty0 (11):\penalty0 e16729, 2019.

\bibitem[Kocijan et~al.(2004)Kocijan, Murray-Smith, Rasmussen, and
  Girard]{Kocijan2004}
Ju{\v{s}} Kocijan, Roderick Murray-Smith, Carl~Edward Rasmussen, and Agathe
  Girard.
\newblock {G}aussian process model based predictive control.
\newblock In \emph{Proceeding of American Control Conference}, volume~3, pages
  2214--2219, 2004.

\bibitem[Hewing et~al.(2020)Hewing, Kabzan, and Zeilinger]{Hewing2020}
L.~Hewing, J.~Kabzan, and M.~N. Zeilinger.
\newblock Cautious model predictive control using {G}aussian process
  regression.
\newblock \emph{IEEE Transactions on Control Systems Technology}, 28\penalty0
  (6):\penalty0 2736--2743, 2020.

\bibitem[Klimasauskas(1998)]{Klimasauskas1998}
Casimir~C. Klimasauskas.
\newblock Hybrid modeling for robust nonlinear multivariable control.
\newblock \emph{ISA Transactions}, 37\penalty0 (4):\penalty0 291--297, 1998.

\bibitem[Zhang et~al.(2019{\natexlab{b}})Zhang, Wu, Rincon, and
  Christofides]{Zhang2019}
Zhihao Zhang, Zhe Wu, David Rincon, and Panagiotis~D. Christofides.
\newblock Real-time optimization and control of nonlinear processes using
  machine learning.
\newblock \emph{Mathematics}, 7\penalty0 (10):\penalty0 890,
  2019{\natexlab{b}}.
\newblock \doi{10.3390/math7100890}.

\bibitem[Teixeira et~al.(2006)Teixeira, Clemente, Cunha, Carrondo, and
  Oliveira]{Teixeira2006}
Ana~P. Teixeira, Jo\~ao~J. Clemente, Ant\'onio~E. Cunha, Manuel J.~T. Carrondo,
  and Rui Oliveira.
\newblock Bioprocess iterative batch-to-batch optimization based on hybrid
  parametric/nonparametric models.
\newblock \emph{Biotechnology Progress}, 22\penalty0 (1):\penalty0 247--258,
  2006.

\bibitem[Cubillos et~al.(2007)Cubillos, Acu\~na, and Lima]{Cubillos2007}
F.~A. Cubillos, G.~Acu\~na, and E.L. Lima.
\newblock Real-time process optimization based on grey-box neural models.
\newblock \emph{Brazilian Journal of Chemical Engineering}, 24:\penalty0
  433--443, 2007.

\bibitem[Chachuat et~al.(2009)Chachuat, Srinivasan, and Bonvin]{Chachuat2009}
B.~Chachuat, B.~Srinivasan, and D.~Bonvin.
\newblock Adaptation strategies for real-time optimization.
\newblock \emph{Computers {\&} Chemical Engineering}, 33\penalty0
  (10):\penalty0 1557--1567, 2009.

\bibitem[Marchetti et~al.(2009)Marchetti, Chachuat, and Bonvin]{Marchetti2009}
A.~Marchetti, B.~Chachuat, and D.~Bonvin.
\newblock Modifier-adaptation methodology for real-time optimization.
\newblock \emph{Industrial \& Engineering Chemistry Research}, 48\penalty0
  (13):\penalty0 6022--6033, 2009.

\bibitem[Gao et~al.(2016)Gao, Wenzel, and Engell]{Gao2016}
Weihua Gao, Simon Wenzel, and Sebastian Engell.
\newblock A reliable modifier-adaptation strategy for real-time optimization.
\newblock \emph{Computers \& Chemical Engineering}, 91:\penalty0 318--328,
  2016.

\bibitem[Singhal et~al.(2016)Singhal, Marchetti, Faulwasser, and
  Bonvin]{Singhal2016}
Martand Singhal, Alejandro~G. Marchetti, Timm Faulwasser, and Dominique Bonvin.
\newblock Real-time optimization based on adaptation of surrogate models.
\newblock \emph{IFAC-PapersOnLine}, 49\penalty0 (7):\penalty0 412--417, 2016.

\bibitem[Milosavljevic et~al.(2020)Milosavljevic, Marchetti, Cortinovis,
  Faulwasser, Mercangöz, and Bonvin]{Milosavljevic2020}
P.~Milosavljevic, A.~G. Marchetti, A.~Cortinovis, T.~Faulwasser, M.~Mercangöz,
  and D.~Bonvin.
\newblock Real-time optimization of load sharing for gas compressors in the
  presence of uncertainty.
\newblock \emph{Applied Energy}, 272:\penalty0 114883, 2020.

\bibitem[de~Avila~Ferreira et~al.(2019)de~Avila~Ferreira, Wuillemin, Marchetti,
  Salzmann, {Van Herle}, and Bonvin]{deAvila2019}
T.~de~Avila~Ferreira, Z.~Wuillemin, A.G. Marchetti, C.~Salzmann, J.~{Van
  Herle}, and D.~Bonvin.
\newblock Real-time optimization of an experimental solid-oxide fuel-cell
  system.
\newblock \emph{Journal of Power Sources}, 429:\penalty0 168--179, 2019.

\bibitem[Ferreira et~al.(2018)Ferreira, Shukla, Faulwasser, Jones, and
  Bonvin]{Ferreira2018}
T.~d.~A. Ferreira, H.~A. Shukla, T.~Faulwasser, C.~N. Jones, and D.~Bonvin.
\newblock Real-time optimization of uncertain process systems via modifier
  adaptation and {G}aussian processes.
\newblock In \emph{2018 European Control Conference (ECC)}, pages 465--470,
  2018.

\bibitem[del Rio~Chanona et~al.(2019)del Rio~Chanona, Graciano, Bradford, and
  Chachuat]{delRio2019}
Ehecatl~Antonio del Rio~Chanona, JE~Alves Graciano, Eric Bradford, and Benoit
  Chachuat.
\newblock Modifier-adaptation schemes employing {G}aussian processes and trust
  regions for real-time optimization.
\newblock \emph{IFAC-PapersOnLine}, 52\penalty0 (1):\penalty0 52--57, 2019.

\bibitem[del Rio~Chanona et~al.(2021)del Rio~Chanona, Petsagkourakis, Bradford,
  Graciano, and Chachuat]{delRio2021}
Ehecatl~Antonio del Rio~Chanona, Panagiotis Petsagkourakis, Eric Bradford,
  JE~Alves Graciano, and Beno{\^\i}t Chachuat.
\newblock Real-time optimization meets {B}ayesian optimization and
  derivative-free optimization: A tale of modifier adaptation.
\newblock \emph{Computers \& Chemical Engineering}, 147:\penalty0 107249, 2021.

\bibitem[Petsagkourakis et~al.(2021)Petsagkourakis, Chachuat, and del
  Rio-Chanona]{Petsagkourakis2021}
P.~Petsagkourakis, B.~Chachuat, and E.~A. del Rio-Chanona.
\newblock Safe real-time optimization using multi-fidelity {G}aussian
  processes.
\newblock In \emph{60th IEEE Conference on Decision and Control (CDC)}, pages
  6734--6741, 2021.

\bibitem[Sutton and Barto(2018)]{Sutton2018}
Richard~S Sutton and Andrew~G Barto.
\newblock \emph{{Reinforcement learning: An introduction}}.
\newblock MIT press, 2018.

\bibitem[Bertsekas(2012)]{bertsekas2012dynamic}
Dimitri Bertsekas.
\newblock \emph{Dynamic programming and optimal control: Volume I}, volume~1.
\newblock Athena scientific, 2012.

\bibitem[Levine et~al.(2016)Levine, Finn, Darrell, and Abbeel]{Levine2016}
Sergey Levine, Chelsea Finn, Trevor Darrell, and Pieter Abbeel.
\newblock {End-to-end training of deep visuomotor policies}.
\newblock \emph{The Journal of Machine Learning Research}, 17\penalty0
  (1):\penalty0 1334--1373, 2016.

\bibitem[Williams et~al.(2016)Williams, Drews, Goldfain, Rehg, and
  Theodorou]{williams2016aggressive}
Grady Williams, Paul Drews, Brian Goldfain, James~M Rehg, and Evangelos~A
  Theodorou.
\newblock Aggressive driving with model predictive path integral control.
\newblock In \emph{IEEE International Conference on Robotics and Automation
  (ICRA)}, pages 1433--1440, 2016.

\bibitem[Silver et~al.(2017)Silver, Schrittwieser, Simonyan, Antonoglou, Huang,
  Guez, Hubert, Baker, Lai, Bolton, et~al.]{Silver2017}
David Silver, Julian Schrittwieser, Karen Simonyan, Ioannis Antonoglou, Aja
  Huang, Arthur Guez, Thomas Hubert, Lucas Baker, Matthew Lai, Adrian Bolton,
  et~al.
\newblock {Mastering the game of Go without human knowledge}.
\newblock \emph{Nature}, 550\penalty0 (7676):\penalty0 354, 2017.

\bibitem[Bu{\c{s}}oniu et~al.(2018)Bu{\c{s}}oniu, de~Bruin, Toli{\'c}, Kober,
  and Palunko]{bucsoniu2018reinforcement}
Lucian Bu{\c{s}}oniu, Tim de~Bruin, Domagoj Toli{\'c}, Jens Kober, and Ivana
  Palunko.
\newblock {Reinforcement learning for control: Performance, stability, and deep
  approximators}.
\newblock \emph{Annual Reviews in Control}, pages 8--28, 2018.

\bibitem[Yoo et~al.(2021{\natexlab{a}})Yoo, Byun, Han, and Lee]{yoo2021review}
Haeun Yoo, Ha~Eun Byun, Dongho Han, and Jay~H Lee.
\newblock Reinforcement learning for batch process control: Review and
  perspectives.
\newblock \emph{Annual Reviews in Control}, 52:\penalty0 108--119,
  2021{\natexlab{a}}.

\bibitem[Wilson and Martinez(1997)]{wilson1997neuro}
JA~Wilson and EC~Martinez.
\newblock Neuro-fuzzy modeling and control of a batch process involving
  simultaneous reaction and distillation.
\newblock \emph{Computers \& Chemical Engineering}, 21:\penalty0 S1233--S1238,
  1997.

\bibitem[Kaisare et~al.(2003)Kaisare, Lee, and Lee]{kaisare2003simulation}
Niket~S Kaisare, Jong~Min Lee, and Jay~H Lee.
\newblock Simulation based strategy for nonlinear optimal control: Application
  to a microbial cell reactor.
\newblock \emph{International Journal of Robust and Nonlinear Control:
  IFAC-Affiliated Journal}, 13\penalty0 (3-4):\penalty0 347--363, 2003.

\bibitem[Peroni et~al.(2005)Peroni, Kaisare, and Lee]{peroni2005optimal}
Catalina~Valencia Peroni, Niket~S Kaisare, and Jay~H Lee.
\newblock Optimal control of a fed-batch bioreactor using simulation-based
  approximate dynamic programming.
\newblock \emph{IEEE Transactions on Control Systems Technology}, 13\penalty0
  (5):\penalty0 786--790, 2005.

\bibitem[Lee et~al.(2006)Lee, Kaisare, and Lee]{lee2006choice}
Jong~Min Lee, Niket~S Kaisare, and Jay~H Lee.
\newblock {Choice of approximator and design of penalty function for an
  approximate dynamic programming based control approach}.
\newblock \emph{Journal of Process Control}, 16\penalty0 (2):\penalty0
  135--156, 2006.

\bibitem[Lee and Lee(2009)]{lee2009approximate}
Jong~Min Lee and Jay~H Lee.
\newblock An approximate dynamic programming based approach to dual adaptive
  control.
\newblock \emph{Journal of process control}, 19\penalty0 (5):\penalty0
  859--864, 2009.

\bibitem[Nosair et~al.(2010)Nosair, Yang, and Lee]{nosair2010min}
Hussam Nosair, Yu~Yang, and Jong~Min Lee.
\newblock {Min--max control using parametric approximate dynamic programming}.
\newblock \emph{Control Engineering Practice}, 18\penalty0 (2):\penalty0
  190--197, 2010.

\bibitem[Yang and Lee(2013)]{yang2013switching}
Yu~Yang and Jong~Min Lee.
\newblock A switching robust model predictive control approach for nonlinear
  systems.
\newblock \emph{Journal of Process Control}, 23\penalty0 (6):\penalty0
  852--860, 2013.

\bibitem[Zhu et~al.(2020)Zhu, Cui, Takami, Kanokogi, and
  Matsubara]{zhu2020scalable}
Lingwei Zhu, Yunduan Cui, Go~Takami, Hiroaki Kanokogi, and Takamitsu Matsubara.
\newblock Scalable reinforcement learning for plant-wide control of vinyl
  acetate monomer process.
\newblock \emph{Control Engineering Practice}, 97:\penalty0 104331, 2020.

\bibitem[Sun et~al.(2018)Sun, He, Wang, Gui, Yang, and Zhu]{sun2018data}
Bei Sun, Mingfang He, Yalin Wang, Weihua Gui, Chunhua Yang, and Quanmin Zhu.
\newblock A data-driven optimal control approach for solution purification
  process.
\newblock \emph{Journal of Process Control}, 68:\penalty0 171--185, 2018.

\bibitem[Ge et~al.(2018)Ge, Li, and Chang]{ge2018approximate}
Yulei Ge, Shurong Li, and Peng Chang.
\newblock {An approximate dynamic programming method for the optimal control of
  Alkai-Surfactant-Polymer flooding}.
\newblock \emph{Journal of Process Control}, 64:\penalty0 15--26, 2018.

\bibitem[Kim et~al.(2018)Kim, Choi, and Lee]{kim2018pomdp}
Jong~Woo Kim, Go~Bong Choi, and Jong~Min Lee.
\newblock {A POMDP framework for integrated scheduling of infrastructure
  maintenance and inspection}.
\newblock \emph{Computers \& Chemical Engineering}, 112:\penalty0 239--252,
  2018.

\bibitem[Pandian and Noel(2018)]{pandian2018control}
B~Jaganatha Pandian and Mathew~Mithra Noel.
\newblock Control of a bioreactor using a new partially supervised
  reinforcement learning algorithm.
\newblock \emph{Journal of Process Control}, 69:\penalty0 16--29, 2018.

\bibitem[Mnih et~al.(2015)Mnih, Kavukcuoglu, Silver, Rusu, Veness, Bellemare,
  Graves, Riedmiller, Fidjeland, and Ostrovski]{Mnih2015}
Volodymyr Mnih, Koray Kavukcuoglu, David Silver, Andrei~A Rusu, Joel Veness,
  Marc~G Bellemare, Alex Graves, Martin Riedmiller, Andreas~K Fidjeland, and
  Georg Ostrovski.
\newblock {Human--level control through deep reinforcement learning}.
\newblock \emph{Nature}, 518\penalty0 (7540):\penalty0 529, 2015.

\bibitem[Wang et~al.(2018{\natexlab{b}})Wang, Velswamy, and
  Huang]{wang2018NovelApproach}
Yuan Wang, Kirubakaran Velswamy, and Biao Huang.
\newblock A novel approach to feedback control with deep reinforcement
  learning.
\newblock \emph{IFAC-PapersOnLine}, 51\penalty0 (18):\penalty0 31--36,
  2018{\natexlab{b}}.

\bibitem[Schulman et~al.(2017)Schulman, Wolski, Dhariwal, Radford, and
  Klimov]{schulman2017ProximalPolicy}
John Schulman, Filip Wolski, Prafulla Dhariwal, Alec Radford, and Oleg Klimov.
\newblock Proximal policy optimization algorithms.
\newblock \emph{arXiv preprint arXiv:1707.06347}, 2017.

\bibitem[Ma et~al.(2019)Ma, Zhu, Benton, and Romagnoli]{ma2019continuous}
Yan Ma, Wenbo Zhu, Michael~G Benton, and Jos{\'e} Romagnoli.
\newblock Continuous control of a polymerization system with deep reinforcement
  learning.
\newblock \emph{Journal of Process Control}, 75:\penalty0 40--47, 2019.

\bibitem[Lillicrap et~al.(2015)Lillicrap, Hunt, Pritzel, Heess, Erez, Tassa,
  Silver, and Wierstra]{lillicrap2015continuous}
Timothy~P Lillicrap, Jonathan~J Hunt, Alexander Pritzel, Nicolas Heess, Tom
  Erez, Yuval Tassa, David Silver, and Daan Wierstra.
\newblock Continuous control with deep reinforcement learning.
\newblock \emph{arXiv preprint arXiv:1509.02971}, 2015.

\bibitem[Oh et~al.(2021)Oh, Kim, Son, Kim, Lee, and Lee]{oh2021automatic}
Tae~Hoon Oh, Jong~Woo Kim, Sang~Hwan Son, Hosoo Kim, Kyungmoo Lee, and Jong~Min
  Lee.
\newblock Automatic control of simulated moving bed process with deep
  {Q}-network.
\newblock \emph{Journal of Chromatography A}, 1647:\penalty0 462073, 2021.

\bibitem[Petsagkourakis et~al.(2020)Petsagkourakis, Sandoval, Bradford, Zhang,
  and del Rio-Chanona]{petsagkourakis2020reinforcement}
Panagiotis Petsagkourakis, Ilya~Orson Sandoval, Eric Bradford, Dongda Zhang,
  and Ehecatl~Antonio del Rio-Chanona.
\newblock Reinforcement learning for batch bioprocess optimization.
\newblock \emph{Computers \& Chemical Engineering}, 133:\penalty0 106649, 2020.

\bibitem[Williams(1992)]{williams1992SimpleStatistical}
Ronald~J Williams.
\newblock Simple statistical gradient-following algorithms for connectionist
  reinforcement learning.
\newblock \emph{Machine learning}, 8:\penalty0 229--256, 1992.

\bibitem[Bao et~al.(2021)Bao, Zhu, and Qian]{bao2021DeepReinforcement}
Yaoyao Bao, Yuanming Zhu, and Feng Qian.
\newblock A {{Deep Reinforcement Learning Approach}} to {{Improve}} the
  {{Learning Performance}} in {{Process Control}}.
\newblock \emph{Industrial \& Engineering Chemistry Research}, page
  acs.iecr.0c05678, 2021.

\bibitem[Fujimoto et~al.(2018)Fujimoto, van Hoof, and
  Meger]{fujimoto2018addressing}
Scott Fujimoto, Herke van Hoof, and David Meger.
\newblock Addressing function approximation error in actor-critic methods.
\newblock In \emph{Proceedings of the 35th International Conference on Machine
  Learning}, volume~80 of \emph{Proceedings of Machine Learning Research},
  pages 1587--1596. PMLR, 10--15 Jul 2018.

\bibitem[Dogru et~al.(2021)Dogru, Wieczorek, Velswamy, Ibrahim, and
  Huang]{dogru2021online}
Oguzhan Dogru, Nathan Wieczorek, Kirubakaran Velswamy, Fadi Ibrahim, and Biao
  Huang.
\newblock Online reinforcement learning for a continuous space system with
  experimental validation.
\newblock \emph{Journal of Process Control}, 104:\penalty0 86--100, 2021.

\bibitem[Mnih et~al.(2016)Mnih, Badia, Mirza, Graves, Lillicrap, Harley,
  Silver, and Kavukcuoglu]{mnih2016AsynchronousMethods}
Volodymyr Mnih, Adria~Puigdomenech Badia, Mehdi Mirza, Alex Graves, Timothy
  Lillicrap, Tim Harley, David Silver, and Koray Kavukcuoglu.
\newblock Asynchronous methods for deep reinforcement learning.
\newblock In \emph{International conference on machine learning}, pages
  1928--1937. PMLR, 2016.

\bibitem[Joshi et~al.(2021)Joshi, Makker, Kodamana, and Kandath]{joshi2021twin}
Tanuja Joshi, Shikhar Makker, Hariprasad Kodamana, and Harikumar Kandath.
\newblock {Twin actor twin delayed deep deterministic policy gradient (TATD3)
  learning for batch process control}.
\newblock \emph{Computers \& Chemical Engineering}, 155:\penalty0 107527, 2021.

\bibitem[Mowbray et~al.(2021)Mowbray, Smith, {Del Rio-Chanona}, and
  Zhang]{mowbray2021UsingProcess}
Max Mowbray, Robin Smith, Ehecatl~A {Del Rio-Chanona}, and Dongda Zhang.
\newblock Using process data to generate an optimal control policy via
  apprenticeship and reinforcement learning.
\newblock \emph{AIChE Journal}, page e17306, 2021.

\bibitem[Yoo et~al.(2021{\natexlab{b}})Yoo, Kim, Kim, and
  Lee]{yoo2021reinforcement}
Haeun Yoo, Boeun Kim, Jong~Woo Kim, and Jay~H Lee.
\newblock {Reinforcement learning based optimal control of batch processes
  using Monte-Carlo deep deterministic policy gradient with phase
  segmentation}.
\newblock \emph{Computers \& Chemical Engineering}, 144:\penalty0 107133,
  2021{\natexlab{b}}.

\bibitem[Lawrence et~al.(2022)Lawrence, Forbes, Loewen, McClement,
  Backstr{\"o}m, and Gopaluni]{lawrence2022deep}
Nathan~P. Lawrence, Michael~G. Forbes, Philip~D. Loewen, Daniel~G. McClement,
  Johan~U. Backstr{\"o}m, and R.~Bhushan Gopaluni.
\newblock Deep reinforcement learning with shallow controllers: An experimental
  application to {PID} tuning.
\newblock \emph{Control Engineering Practice}, 121:\penalty0 105046, 2022.

\bibitem[Zhu et~al.(2022)Zhu, Castillo, Wang, Rendall, Chiang, Hayot, and
  Romagnoli]{zhu2022benchmark}
Wenbo Zhu, Ivan Castillo, Zhenyu Wang, Ricardo Rendall, Leo~H Chiang, Philippe
  Hayot, and Jose~A Romagnoli.
\newblock Benchmark study of reinforcement learning in controlling and
  optimizing batch processes.
\newblock \emph{Journal of Advanced Manufacturing and Processing}, 4\penalty0
  (2):\penalty0 e10113, 2022.

\bibitem[Haarnoja et~al.(2018)Haarnoja, Zhou, Abbeel, and
  Levine]{haarnoja2018soft}
Tuomas Haarnoja, Aurick Zhou, Pieter Abbeel, and Sergey Levine.
\newblock Soft actor-critic: Off-policy maximum entropy deep reinforcement
  learning with a stochastic actor.
\newblock In \emph{Proceedings of the 35th International Conference on Machine
  Learning}, volume~80 of \emph{Proceedings of Machine Learning Research},
  pages 1861--1870. PMLR, 10--15 Jul 2018.

\bibitem[Janjua et~al.(2023)Janjua, Shah, White, Miahi, Machado, and
  White]{janjua2023gvfs}
Muhammad~Kamran Janjua, Haseeb Shah, Martha White, Erfan Miahi, Marlos~C
  Machado, and Adam White.
\newblock Gvfs in the real world: making predictions online for water
  treatment.
\newblock \emph{Machine Learning}, pages 1--31, 2023.

\bibitem[Sutton et~al.(2011)Sutton, Modayil, Delp, Degris, Pilarski, White, and
  Precup]{sutton2011horde}
Richard~S Sutton, Joseph Modayil, Michael Delp, Thomas Degris, Patrick~M
  Pilarski, Adam White, and Doina Precup.
\newblock Horde: A scalable real-time architecture for learning knowledge from
  unsupervised sensorimotor interaction.
\newblock In \emph{The 10th International Conference on Autonomous Agents and
  Multiagent Systems-Volume 2}, pages 761--768, 2011.

\bibitem[Sedighizadeh and Rezazadeh(2008)]{sedighizadeh2008AdaptivePID}
M~Sedighizadeh and A~Rezazadeh.
\newblock Adaptive {{PID}} controller based on reinforcement learning for wind
  turbine control.
\newblock In \emph{Proceedings of World Academy of Science, Engineering and
  Technology}, volume~27, pages 257--262. {Citeseer}, 2008.

\bibitem[Carlucho et~al.(2017)Carlucho, De~Paula, Villar, and
  Acosta]{carlucho2017IncrementalLearning}
Ignacio Carlucho, Mariano De~Paula, Sebastian~A. Villar, and Gerardo~G. Acosta.
\newblock Incremental {{Q-learning}} strategy for adaptive {{PID}} control of
  mobile robots.
\newblock \emph{Expert Systems with Applications}, 80:\penalty0 183--199, 2017.

\bibitem[Shipman and Coetzee(2019)]{shipman2019ReinforcementLearning}
William~J. Shipman and Loutjie~C. Coetzee.
\newblock Reinforcement learning and deep neural networks for {{PI}} controller
  tuning.
\newblock \emph{IFAC-PapersOnLine}, 52\penalty0 (14):\penalty0 111--116, 2019.

\bibitem[Kumar and Ramadge(2021)]{kumar2021DiffLoopTuning}
Athindran~Ramesh Kumar and Peter~J. Ramadge.
\newblock {{DiffLoop}}: Tuning {{PID}} controllers by differentiating through
  the feedback loop.
\newblock In \emph{2021 55th {{Annual Conference}} on {{Information Sciences}}
  and {{Systems}} ({{CISS}})}, pages 1--6, 2021.

\bibitem[Lakhani et~al.(2022)Lakhani, Chowdhury, and
  Lu]{lakhani2022StabilitypreservingAutomatic}
Ayub~I. Lakhani, Myisha~A. Chowdhury, and Qiugang Lu.
\newblock Stability-preserving automatic tuning of {PID} control with
  reinforcement learning.
\newblock \emph{Complex Engineering Systems}, 2\penalty0 (1):\penalty0 3, 2022.

\bibitem[Lee and Lee(2006)]{lee2006approximate}
Jay~H Lee and Jong~Min Lee.
\newblock Approximate dynamic programming based approach to process control and
  scheduling.
\newblock \emph{Computers \& Chemical Engineering}, 30\penalty0
  (10-12):\penalty0 1603--1618, 2006.

\bibitem[Brujeni et~al.(2010)Brujeni, Lee, and Shah]{brujeni2010dynamic}
Lena~Abbasi Brujeni, Jong~Min Lee, and Sirish~L Shah.
\newblock \emph{{Dynamic tuning of PI-controllers based on model-free
  reinforcement learning methods}}.
\newblock IEEE, 2010.

\bibitem[Berger and da~Fonseca~Neto(2013)]{berger2013neurodynamic}
Marcus~AR Berger and Jo{\~A}o~Viana da~Fonseca~Neto.
\newblock {Neurodynamic programming approach for the {PID} controller
  adaptation}.
\newblock \emph{IFAC Proceedings Volumes}, 46\penalty0 (11):\penalty0 534--539,
  2013.

\bibitem[Shafi et~al.(2020)Shafi, Velswamy, Ibrahim, and
  Huang]{shafi2020hierarchical}
Hareem Shafi, Kirubakaran Velswamy, Fadi Ibrahim, and Biao Huang.
\newblock {A Hierarchical Constrained Reinforcement Learning for Optimization
  of Bitumen Recovery Rate in a Primary Separation Vessel}.
\newblock \emph{Computers \& Chemical Engineering}, page 106939, 2020.

\bibitem[Kim et~al.(2021)Kim, Park, Oh, and Lee]{kim2021model}
Jong~Woo Kim, Byung~Jun Park, Tae~Hoon Oh, and Jong~Min Lee.
\newblock Model-based reinforcement learning and predictive control for
  two-stage optimal control of fed-batch bioreactor.
\newblock \emph{Computers \& Chemical Engineering}, 154:\penalty0 107465, 2021.

\bibitem[Wang et~al.(2019)Wang, Bao, Clavera, Hoang, Wen, Langlois, Zhang,
  Zhang, Abbeel, and Ba]{wang2019benchmarking}
Tingwu Wang, Xuchan Bao, Ignasi Clavera, Jerrick Hoang, Yeming Wen, Eric
  Langlois, Shunshi Zhang, Guodong Zhang, Pieter Abbeel, and Jimmy Ba.
\newblock {Benchmarking model-based reinforcement learning}.
\newblock \emph{arXiv preprint arXiv:1907.02057}, 2019.

\bibitem[Syafiie et~al.(2011)Syafiie, Tadeo, Martinez, and
  Alvarez]{syafiie2011ModelfreeControl}
S.~Syafiie, F.~Tadeo, E.~Martinez, and T.~Alvarez.
\newblock Model-free control based on reinforcement learning for a wastewater
  treatment problem.
\newblock \emph{Applied Soft Computing}, 11\penalty0 (1):\penalty0 73--82,
  2011.

\bibitem[Henderson et~al.(2018)Henderson, Islam, Bachman, Pineau, Precup, and
  Meger]{henderson2018deep}
Peter Henderson, Riashat Islam, Philip Bachman, Joelle Pineau, %doina Precup,
  and David Meger.
\newblock Deep reinforcement learning that matters.
\newblock In \emph{Thirty-Second AAAI Conference on Artificial Intelligence},
  2018.

\bibitem[Recht(2019)]{recht2019tour}
Benjamin Recht.
\newblock A tour of reinforcement learning: The view from continuous control.
\newblock \emph{Annual Review of Control, Robotics, and Autonomous Systems},
  2:\penalty0 253--279, 2019.

\bibitem[Prokhorov and Wunsch(1997)]{Prokhorov1997ACD}
Danil~V Prokhorov and Donald~C Wunsch.
\newblock Adaptive critic designs.
\newblock \emph{IEEE transactions on Neural Networks}, 8\penalty0 (5):\penalty0
  997--1007, 1997.

\bibitem[Lewis and Vrabie(2009)]{Lewis2009RL}
Frank~L Lewis and Draguna Vrabie.
\newblock Reinforcement learning and adaptive dynamic programming for feedback
  control.
\newblock \emph{{IEEE Circuits and Systems Magazine}}, 9\penalty0 (3), 2009.

\bibitem[Jiang and Jiang(2014)]{jiang2014robust}
Yu~Jiang and Zhong-Ping Jiang.
\newblock Robust adaptive dynamic programming and feedback stabilization of
  nonlinear systems.
\newblock \emph{IEEE Transactions on Neural Networks and Learning Systems},
  25\penalty0 (5):\penalty0 882--893, 2014.

\bibitem[Kim et~al.(2020{\natexlab{a}})Kim, Park, Yoo, Oh, Lee, and
  Lee]{kim2020gdhp}
Jong~Woo Kim, Byung~Jun Park, Haeun Yoo, Tae~Hoon Oh, Jay~H Lee, and Jong~Min
  Lee.
\newblock {A model-based deep reinforcement learning method applied to
  finite-horizon optimal control of nonlinear control-affine system}.
\newblock \emph{Journal of Process Control}, 87:\penalty0 166--178,
  2020{\natexlab{a}}.

\bibitem[Kim et~al.(2020{\natexlab{b}})Kim, Oh, Son, Jeong, and
  Lee]{kim2020convergence}
Jong~Woo Kim, Tae~Hoon Oh, Sang~Hwan Son, Dong~Hwi Jeong, and Jong~Min Lee.
\newblock {Convergence analysis of the deep neural networks based globalized
  dual heuristic programming}.
\newblock \emph{Automatica}, 122:\penalty0 109222, 2020{\natexlab{b}}.

\bibitem[Theodorou et~al.(2010)Theodorou, Buchli, and
  Schaal]{theodorou2010generalized}
Evangelos Theodorou, Jonas Buchli, and Stefan Schaal.
\newblock A generalized path integral control approach to reinforcement
  learning.
\newblock \emph{Journal of machine learning research}, 11\penalty0
  (Nov):\penalty0 3137--3181, 2010.

\bibitem[D'Oro and Ja{\'s}kowski(2020)]{d2020learn}
Pierluca D'Oro and Wojciech Ja{\'s}kowski.
\newblock How to learn a useful critic? {Model}-based action-gradient-estimator
  policy optimization.
\newblock \emph{Advances in Neural Information Processing Systems}, 33, 2020.

\bibitem[Janner et~al.(2019)Janner, Fu, Zhang, and Levine]{janner2019trust}
Michael Janner, Justin Fu, Marvin Zhang, and Sergey Levine.
\newblock When to trust your model: Model-based policy optimization.
\newblock In \emph{Advances in Neural Information Processing Systems}, pages
  12498--12509, 2019.

\bibitem[Levine et~al.(2020)Levine, Kumar, Tucker, and Fu]{levine2020offline}
Sergey Levine, Aviral Kumar, George Tucker, and Justin Fu.
\newblock Offline reinforcement learning: Tutorial, review, and perspectives on
  open problems.
\newblock \emph{arXiv preprint arXiv:2005.01643}, 2020.

\bibitem[Fujimoto et~al.(2019)Fujimoto, Meger, and Precup]{fujimoto2019off}
Scott Fujimoto, David Meger, and Doina Precup.
\newblock Off-policy deep reinforcement learning without exploration.
\newblock In \emph{International Conference on Machine Learning}, pages
  2052--2062, 2019.

\bibitem[Huisman et~al.(2021)Huisman, {van Rijn}, and
  Plaat]{huisman2021SurveyDeep}
Mike Huisman, Jan~N. {van Rijn}, and Aske Plaat.
\newblock A survey of deep meta-learning.
\newblock \emph{Artificial Intelligence Review}, 2021.

\bibitem[Finn et~al.(2017)Finn, Abbeel, and Levine]{finn2017model}
Chelsea Finn, Pieter Abbeel, and Sergey Levine.
\newblock Model-agnostic meta-learning for fast adaptation of deep networks.
\newblock In \emph{Proceedings of the 34th International Conference on Machine
  Learning-Volume 70}, pages 1126--1135, 2017.

\bibitem[Duan et~al.(2016)Duan, Schulman, Chen, Bartlett, Sutskever, and
  Abbeel]{duan2016rl}
Yan Duan, John Schulman, Xi~Chen, Peter~L Bartlett, Ilya Sutskever, and Pieter
  Abbeel.
\newblock $\text{RL}^{2}$: Fast reinforcement learning via slow reinforcement
  learning.
\newblock \emph{arXiv preprint arXiv:1611.02779}, 2016.

\bibitem[McClement et~al.(2022)McClement, Lawrence, Backstr{\"o}m, Loewen,
  Forbes, and Gopaluni]{mcclement2022MetareinforcementLearning}
Daniel~G. McClement, Nathan~P. Lawrence, Johan~U. Backstr{\"o}m, Philip~D.
  Loewen, Michael~G. Forbes, and R.~Bhushan Gopaluni.
\newblock Meta-reinforcement learning for the tuning of {PI} controllers: An
  offline approach.
\newblock \emph{Journal of Process Control}, 118:\penalty0 139--152, 2022.

\bibitem[Masti and Bemporad(2021)]{masti2021learning}
Daniele Masti and Alberto Bemporad.
\newblock Learning nonlinear state--space models using autoencoders.
\newblock \emph{Automatica}, 129:\penalty0 109666, 2021.

\bibitem[Schlegel et~al.(2022)Schlegel, Tkachuk, White, and
  White]{schlegel2022investigating}
Matthew~Kyle Schlegel, Volodymyr Tkachuk, Adam~M White, and Martha White.
\newblock Investigating action encodings in recurrent neural networks in
  reinforcement learning.
\newblock \emph{Transactions on Machine Learning Research}, 2022.

\bibitem[Xie et~al.(2023)Xie, Dogru, Huang, Godwaldt, and
  Willms]{xie2023ReinforcementLearning}
Junyao Xie, Oguzhan Dogru, Biao Huang, Chris Godwaldt, and Brett Willms.
\newblock Reinforcement learning for soft sensor design through autonomous
  cross-domain data selection.
\newblock \emph{Computers \& Chemical Engineering}, 173:\penalty0 108209, 2023.

\bibitem[Esfahani et~al.(2023)Esfahani, Kordabad, Cai, and
  Gros]{nejatbakhshesfahani2023LearningbasedState}
Hossein~Nejatbakhsh Esfahani, Arash~Bahari Kordabad, Wenqi Cai, and Sebastien
  Gros.
\newblock Learning-based state estimation and control using mhe and mpc schemes
  with imperfect models.
\newblock \emph{European Journal of Control}, page 100880, 2023.

\bibitem[Bras et~al.(2023)Bras, Louw, and
  Bradshaw]{bras2023ClassicalActorCritic}
Edward~H. Bras, Tobias~M. Louw, and Steven~M. Bradshaw.
\newblock Classical actor-critic applied to the control of a self-regulatory
  process.
\newblock \emph{IFAC-PapersOnLine}, 56\penalty0 (2):\penalty0 7172--7177, 2023.
\newblock 22nd IFAC World Congress.

\bibitem[Eysenbach et~al.(2022)Eysenbach, Zhang, Levine, and
  Salakhutdinov]{eysenbach2023ContrastiveLearning}
Benjamin Eysenbach, Tianjun Zhang, Sergey Levine, and Russ~R Salakhutdinov.
\newblock Contrastive learning as goal-conditioned reinforcement learning.
\newblock \emph{Advances in Neural Information Processing Systems},
  35:\penalty0 35603--35620, 2022.

\end{thebibliography}





% \end{thebibliography}
\end{document}